%% file: abund2paper.tex
\documentclass{aa}  
\usepackage{natbib}
\usepackage{graphicx}
\usepackage{txfonts}
\begin{document}
   \title{No FIP fractionation in the active stars AR Psc and AY Cet}

   \author{J. Sanz-Forcada\inst{1}
          \and
	  L. Affer\inst{2}
          \and
          G. Micela\inst{2}
          }

   \offprints{J. Sanz-Forcada, \email{jsanz@laeff.inta.es}}

   \institute{Laboratorio de Astrof\'{i}sica Estelar y Exoplanetas,
     Centro de Astrobiolog\'{i}a / CSIC-INTA, LAEFF Campus, P.O. Box 78, 
     E-28691 Villanueva de la
     Ca\~nada, Madrid, Spain\\
     \email{jsanz@laeff.inta.es}
     \and
     INAF -- Osservatorio Astronomico di Palermo
     G. S. Vaiana, Piazza del Parlamento, 1; Palermo, I-90134, Italy\\
     \email{affer@astropa.inaf.it,giusi@astropa.inaf.it}
   }

   \date{Received ; accepted}

 
  \abstract
   {The comparison of coronal and photospheric abundances in cool
  stars is an essential question to resolve. In the Sun an
  enhancement of the elements with low first ionization potential
  (FIP) is observed in the corona with respect to the
  photosphere. Stars with high levels of activity seem to show a
  depletion of elements with low FIP when compared to solar standard
  values; however, the few cases of active stars in which photospheric
  values are available for comparison lead to confusing results, and an
  enlargement of the sample is mandatory for solving this longstanding problem.}
   {We calculate in this paper the photospheric and coronal abundances
  of two well known active binary systems, AR Psc and AY Cet, to get
  further insight into the complications of coronal abundances.}
   {Coronal abundances of 9 elements were calculated by means of the
  reconstruction 
  of a detailed emission measure distribution, using a line-based
  method that considers the lines from different elements
  separately. Photospheric abundances of 8 elements were calculated using
  high-resolution optical spectra of the stars.}
   {The results once again show a lack of any FIP-related effect in
  the coronal abundances of the stars. The presence of metal abundance
  depletion (MAD) or 
  inverse FIP effects in some stars could stem from a mistaken comparison
  to solar photospheric values or from a deficient calculation of
  photospheric abundances in fast-rotating stars.}
   {The lack of FIP fractionation seems to confirm that Alfv\'en waves
  combined with pondermotive forces are dominant in the
  corona of active stars.}


   \maketitle
%

\section{Introduction}

One of the most debated issues in stellar astrophysics is whether the
coronal abundances are similar to the photospheric counterparts in
cool stars. A different composition in the corona and the
photosphere would indicate a physical process taking place
somewhere between the cooler photospheric material and the hotter
corona. Such a process should be capable of distinguishing between
different elements based on a certain physical variable. A
pattern related to that variable would help understanding the
physical processes taking place in the coronal loops.
Such a pattern has been observed in the Sun, which shows, on average,
an enhancement of elements with a low first ionization potential (FIP)
in the corona with respect to the photosphere. The so called ``FIP
effect'' is actually observed in the solar corona and slow wind, but
is absent in coronal holes or fast wind \citep{lam95,feld00}.
Stars similar to the Sun, such as $\alpha$~Cen, show a
similar FIP effect \citep{dra97,raa03}. 
Less active stars, such as Procyon (F4IV),
do not show any fractionation \citep{raa02,sanz04}. 
Intermediate-activity stars,
like $\epsilon$~Eri, 36~Oph or 70~Oph present a much lower FIP effect,
if any is present \citep{lam96,sanz04,woo06,ness08}. 

For the most active stars, 
the results are more uncertain. Their coronal
composition is relatively easy to determine once high spectral
resolution is available. But their high rotation rates
hamper the measurements of photospheric composition due to line
broadening, making the comparison more difficult. 
The photospheric composition can only be well
calculated for stars with low {\it projected} rotational
velocity ($v \sin i$).
Initial studies of
active stars made with XMM-Newton and Chandra high-resolution spectra
found a very different pattern from the Sun \citep[e.g.][and references
  therein]{gud04}. The elements with low FIP would 
actually be underabundant in the corona, a ``metal abundance
depletion'' \citep[``MAD syndrome'',][]{sch96}, or alternatively, 
the elements with high FIP would
be enhanced in the corona \citep[the ``inverse FIP effect'',][]{bri01}. 
A caveat of many of the active stars is that 
their coronal abundances are actually compared to the solar photosphere,
which at least yields to risky conclusions \citep{fav03}. Moreover, most
active stars with photospheric abundances calculated have large
$v \sin i$, with broad photospheric lines. 
As a result, their abundances determination could actually
carry some hidden errors not assessed well by formal
calculations. This is the case for II Peg, AB Dor, or AR Lac (see
Table~\ref{tab:pastabundances}).

\begin{table*}
\caption{Fractionation effects in other stars with known
  photospheric and coronal abundances}\label{tab:pastabundances} 
\tabcolsep 3.pt
\begin{center}
\begin{footnotesize}
 \begin{tabular}{lrcccccc}
\hline \hline
{Star} & HD & Sp. Type & {$L_{\rm X}$} & $L_{\rm X}$/$L_{\rm bol}$  &
$v \sin i$ & FIP effect? & Reference$^a$ \\
       &  & &  (erg s$^{-1}$) &    &  km\,s$^{-1}$ & & \\
\hline
{Sun}        & \dots & G2V & $\sim$27.5  & -6.1  & \dots & FIP effect & FE00 \\
{Procyon}      & 61421 & F4IV & 27.9  & -6.5 & 6.1 & No FIP effect & SF04,RA02 \\
{$\epsilon$ Eri} & 22049 & K2V & 28.2 & -5.1 & 2.4 & Small/No FIP effect & SF04 \\ 
{$\xi$ UMa B}    & 98230 & G5V/[KV]   & 29.5 & -4.3 & 2.8 & (No FIP effect) & BA05 \\
{$\lambda$ And} & 222107 & G8III/? & 30.4 & -4.5 & 7.3 & No FIP effect & SF04 \\ 
{Capella}       &  34029 & G1III/G8III & 30.5 &  -5.3  & 32.7 & (No FIP effect) & BR00 \\
{II Peg}        & 224085 & K2IV/M0-3V & 31.1 &  -2.8   & 21 & Small inverse FIP & HU01  \\
{AB Dor}        &  25647 & K1IV       & 30.1 &  -3.0 & 90 & Inverse FIP & SF03 \\
{AR Lac}       & 210334 & G2IV/K0IV  & 30.9 &  -3.4  & 46/81 & Small inverse FIP & HU03  \\
{V851 Cen}      & 119285 & K2IV-III/? & 30.8 & -3.5 & 6.5 & No FIP effect & SF04 \\
AR Psc        &   8357 & G7V/K1IV & 30.7 & -3.3 & 7.0 & No FIP effect &This work \\
AY Cet        &   7672 & WD/G5III & 31.0 & -4.2 & 4.6 & No FIP effect & This work \\
\hline
\end{tabular}
\end{footnotesize}
\end{center}
{\it Note}: $L_{\rm X}$ (erg s$^{-1}$) calculated in the range
5--100~\AA\ (0.12--2.4 keV).\\
{\it $^a$References for FIP effect}: BA05 \citep{bal05}, BR00 \citep{bri00}, FE00
\citep{feld00}, HU01 \citep{hue01}, HU03 \citep{hue03}, RA02
\citep{raa02}, SF04 \citep{sanz04}, SF03 \citep{sanz03b}. 
\end{table*}

\begin{figure*}
   \centering
   \includegraphics[angle=90,width=0.45\textwidth]{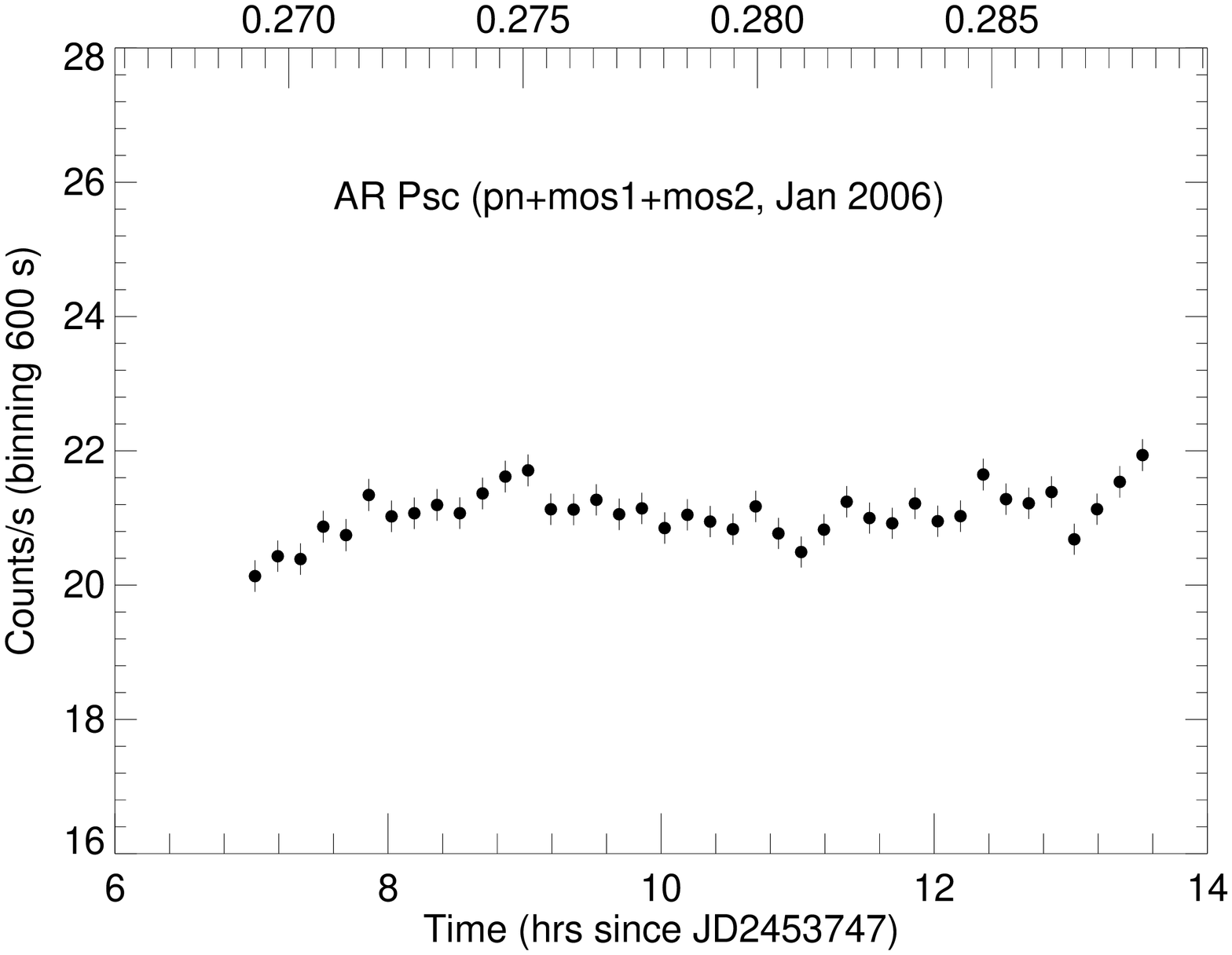}
   \includegraphics[angle=90,width=0.45\textwidth]{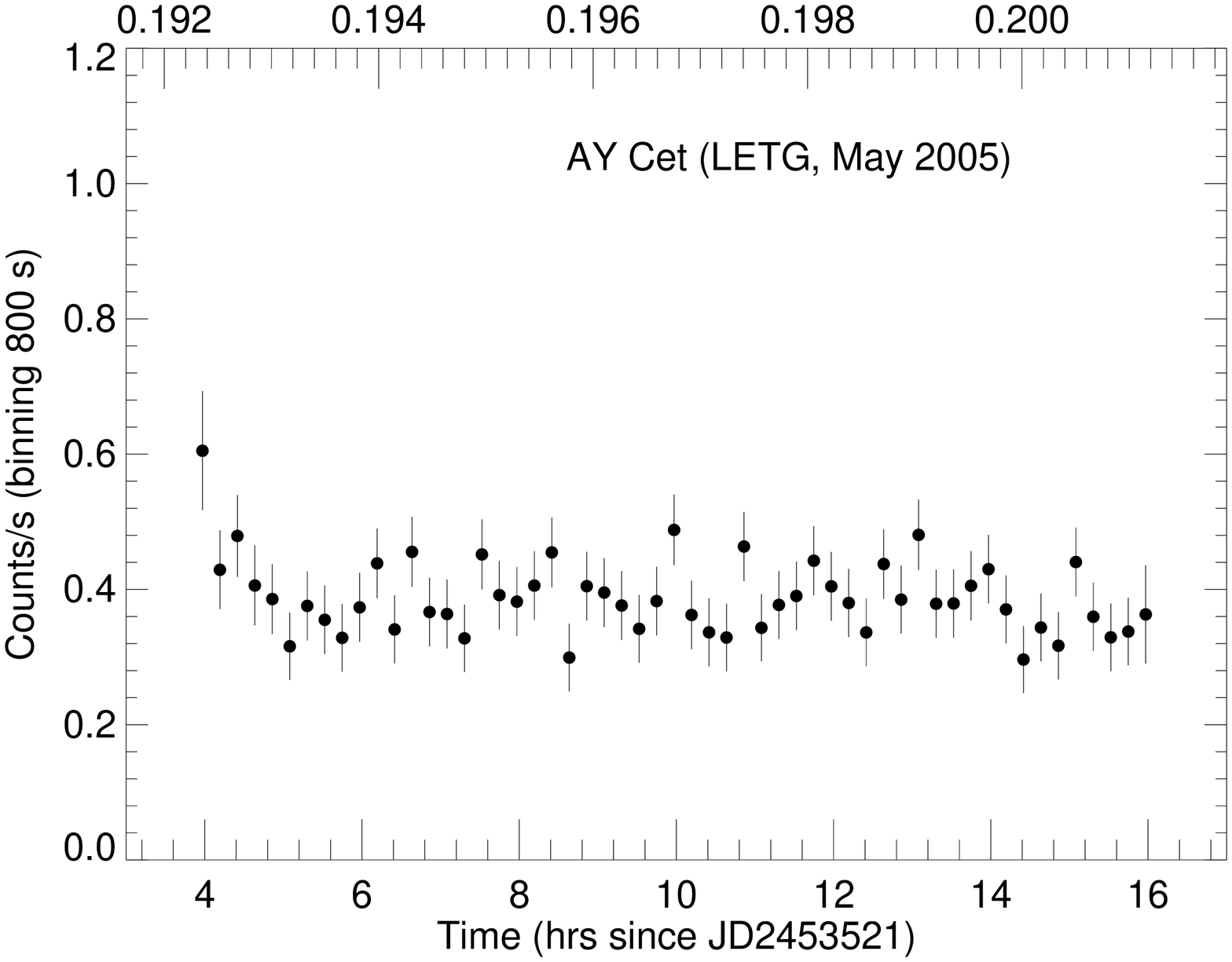}
   \caption{Light curves of AR Psc (using all EPIC detectors) and AY
     Cet (using order +1 and -1 of LETGS) in X-rays. Upper axis
     indicate the orbital phase, assuming that secondary star is
     located behind the primary star at phase 0.}
   \label{fig:lcs}
\end{figure*}

Theoretical explanations of the FIP effect are found in
\citet{lam04}. The authors also propose that Alfv\'en waves,
combined with pondermotive forces, would explain the observed FIP
effect on the Sun, pointing towards the disappearance of any
fractionation of stars with lower or higher activity levels. The same
mechanism could also explain the inverse FIP effect, if such exists,
by reflecting the Alfv\'en waves in the chromosphere. Alf\'en waves
are also being suggested in recent years as responsible for the
energy transportation between the outer convective layers of the star
and the much hotter corona \citep[e.g.][and references
  therein]{erd07,dep07}, one of the most important
problems unresolved in stellar astrophysics. Both questions, the energy
transportation and the FIP fractionation, could actually be connected.

It is thus essential to understand whether active stars suffer an
inverse FIP effect or just no significant fractionation with respect
to the photospheric composition. Given the problems measuring the
photospheric abundances, we should mainly trust
those results for stars with low $v \sin i$. \citet{sanz04}
showed the case of two active stars, $\lambda$~And and V851~Cen,
for which coronal and photospheric abundances are consistent, and no
effect related to FIP would be present. Stars such as Capella
\citep[which shows solar photospheric abundances,][]{bri00,arg03,aud03}
show no inverse FIP effect either, although their
photospheric abundance is poorly known
\citep[{[Fe/H]=--0.16},][]{mcw90}. Other cases of active stars
with no sign of an inverse FIP effect, which are compared to poorly
known photospheric iron abundances, are $\sigma^2$~CrB
\citep{suh05}, EK Dra \citep{tel05}, and YY Men
\citep{aud04}\footnote{Although the authors claim that coronal iron
  is depleted in YY Men, the values in the corona and photosphere of the
  star are actually consistent, once reasonable uncertainties of 
  0.2 dex are considered for the photospheric iron, calculated from a
  low-resolution spectrum by \citet{ran93}.}. 
The present situation is still confusing given the few cases described
in the literature with both photospheric and coronal abundances 
well-calculated. In this 
work we present the results for two more RS~CVn systems known to have
narrow photospheric lines despite their high rotation and activity
level.

\begin{figure*}
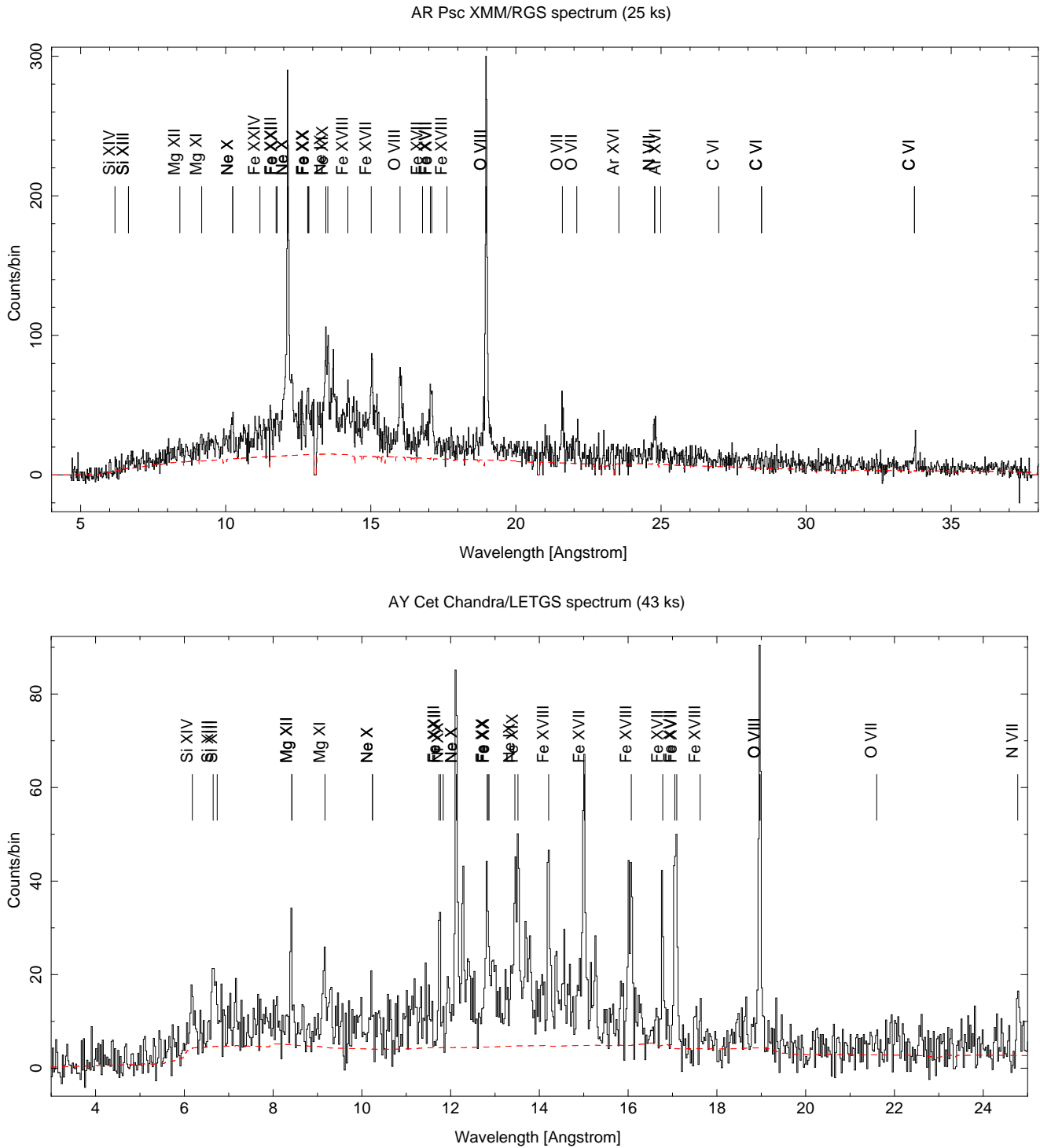

   \centering
   \includegraphics[angle=270,width=0.95\textwidth]{12069f2a.ps}

   \hspace{5mm}

   \includegraphics[angle=270,width=0.95\textwidth]{12069f2b.ps}
   \caption{AR Psc and AY Cet X-ray spectra. The dashed line
     represents the continuum predicted by the EMD.} 
   \label{fig:speccorona}
\end{figure*}

Following \citet{sanz04} we have chosen two active stars that are observed
with their pole on, and therefore have low $v \sin i$.
Thus, their optical spectra display narrow
lines, allowing us to measure the photospheric abundances better. The
two RS CVn binaries are well known X-ray emitters. Their emission
is attributed to coronal activity, powered by the fast rotation forced
by the close binariety. {AR Psc} \citep[G7V/K1IV, $v \sin i$=7
km\,s$^{-1}$,][]{nor04}
and {AY Cet} \citep[WD/G5III, $v \sin i$=4.6 km\,s$^{-1}$,][]{mas08} 
were observed with the
Extreme Ultraviolet Explorer Observatory (EUVE), giving a
first glimpse of its coronal emission \citep{sanz03}.
Optical spectra also indicate a high level of chromospheric activity
\citep{mon97}. Their EMD is similar to other active
stars, either showing an inverse FIP effect, like AB~Dor
\citep{sanz03b,gar05,gar08}, or no
FIP-related fractionation, such as $\lambda$~And \citep{sanz04}. In the case of
AY~Cet, it is not expected that its white dwarf
companion contributes substantially to the X-rays band.
Photospheric abundances were calculated by \citet{ott98} for AY Cet in
only three elements: [Fe/H]=-0.32, [Mg/H]=-0.22, and [Si/H]=-0.32,
with \citet{asp05} values as reference. \citet{sha06} finds a low iron
abundance in AR~Psc, but no information is provided on the reference
system used in the calculations.

The paper is divided as follows: in Sect. 2 the observations are
described. Results are given in Sect. 3, with discussion in Sect. 4, to
finish with the conclusions.

\section{Observations}
Time was awarded (P.I. J. Sanz-Forcada) for observing high-resolution 
X-rays spectra of AY~Cet and AR~Psc.
The Chandra Low Energy Transmission Grating Spectrograph (LETGS)
\citep[$\lambda\lambda\sim$3--175,
$\lambda/\Delta\lambda\sim$60-1000,][]{wei02} observed AY~Cet in May
2005 in combination
with the High Resolution Camera (HRC-S). Data were reduced using the CIAO v4.0
package. The positive and negative orders were summed for the flux
measurements. Lines formed in
the first dispersion order, but contaminated with contribution from
higher dispersion orders, were not employed in the analysis. Light
curves were obtained from the LETG spectra (first and higher 
orders) of AY~Cet with the background properly subtracted
(Fig.~\ref{fig:lcs}).
XMM-Newton observed AR~Psc on 11 January 2006. XMM-Newton observes
simultaneously with the RGS \citep[Reflection Grating
Spectrometer,][]{denher01} ($\lambda\lambda\sim$6--38~\AA,
$\lambda$/$\Delta\lambda\sim$100--500) and the EPIC (European Imaging
Photon Camera) PN and MOS detectors (sensitivity range 0.15--15 keV
and 0.2--10~keV, respectively).
The RGS data were reduced using the standard SAS (Science Analysis
Software) version 8.0.1 package, and the RGS 1 and 2 spectra were
combined to improve the measurement of the line fluxes
(Fig.~\ref{fig:speccorona}). The 
EPIC light curve (Fig.~\ref{fig:lcs}) was constructed by combining the
PN and MOS count rates in a circle around the target, with background
regions properly subtracted using the SAS task {\em epiclccorr}, which
also corrects from several instrumental effects. 

\begin{figure}
   \centering
   \includegraphics[width=0.45\textwidth]{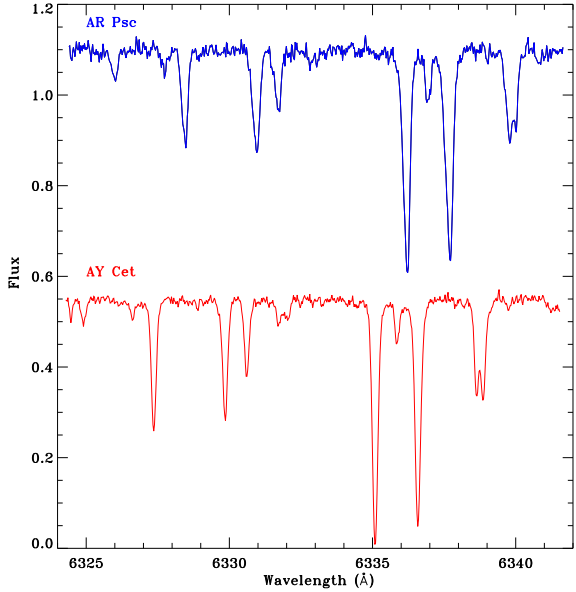}
   \caption{AR Psc and AY Cet optical spectra.} 
   \label{fig:specphot}
\end{figure}

Optical spectra (Fig.~\ref{fig:specphot}) were acquired on 27 and 28 
November 2002 to
calculate the photospheric abundances of the two stars, as part of a
wider observational campaign. The detailed instrumental setup and
observations are described well in \citet{aff05}, so only a short
description is given here. We used the high-resolution cross-dispersed echelle
spectrograph SOFIN, mounted on the Cassegrain focus of the 2.56 m Nordic Optical
Telescope (NOT) located at the Observatorio del Roque de Los Muchachos (La
Palma, Spain). Exposure times ranged from 4 to 20 min,
resulting in high S/N per pixel ($\approx 0.025$ \AA/px)\,
averaging at about 280. A
spectrum of a Th-Ar lamp was obtained following each stellar spectrum, ensuring
accurate wavelength calibration. 
The total spectral range is 3900--9900 \AA, and the resolving power 
is $R=\lambda/\Delta\lambda\,\approx\,80\,000$. The spectra were
reduced with the standard software available 
within the CCDRED and ECHELLE packages of IRAF.\footnote{IRAF (Image
  Reduction and Analysis Facility) is 
distributed by National Optical Astronomy Observatories, operated by the
Association of Universities for Research in Astronomy, Inc., under cooperative
agreement with the National Science Foundation.} The analysis includes overscan
subtraction, flat-fielding, removal of scattered light, extraction of
one-dimensional spectra, wavelength calibration, and continuum
normalization \citep[see][for further details]{aff05}. The
$EW$s were measured using the SPLOT task in
IRAF, assuming a Gaussian profile for weak or moderately strong lines ($EW \la
100$ m\AA) and a Voigt profile for stronger lines. The accuracy (absolute
error) is harder to assess. It almost certainly contains a systematic error due
to the continuum location, because of the presence of interference
fringes (which could not be completely removed) in the redder part of
the stellar spectra, which cause a modulation of 
the local continuum. This error could be particularly important for
the weak lines.

\begin{figure*}
   \centering
   \includegraphics[width=0.498\textwidth]{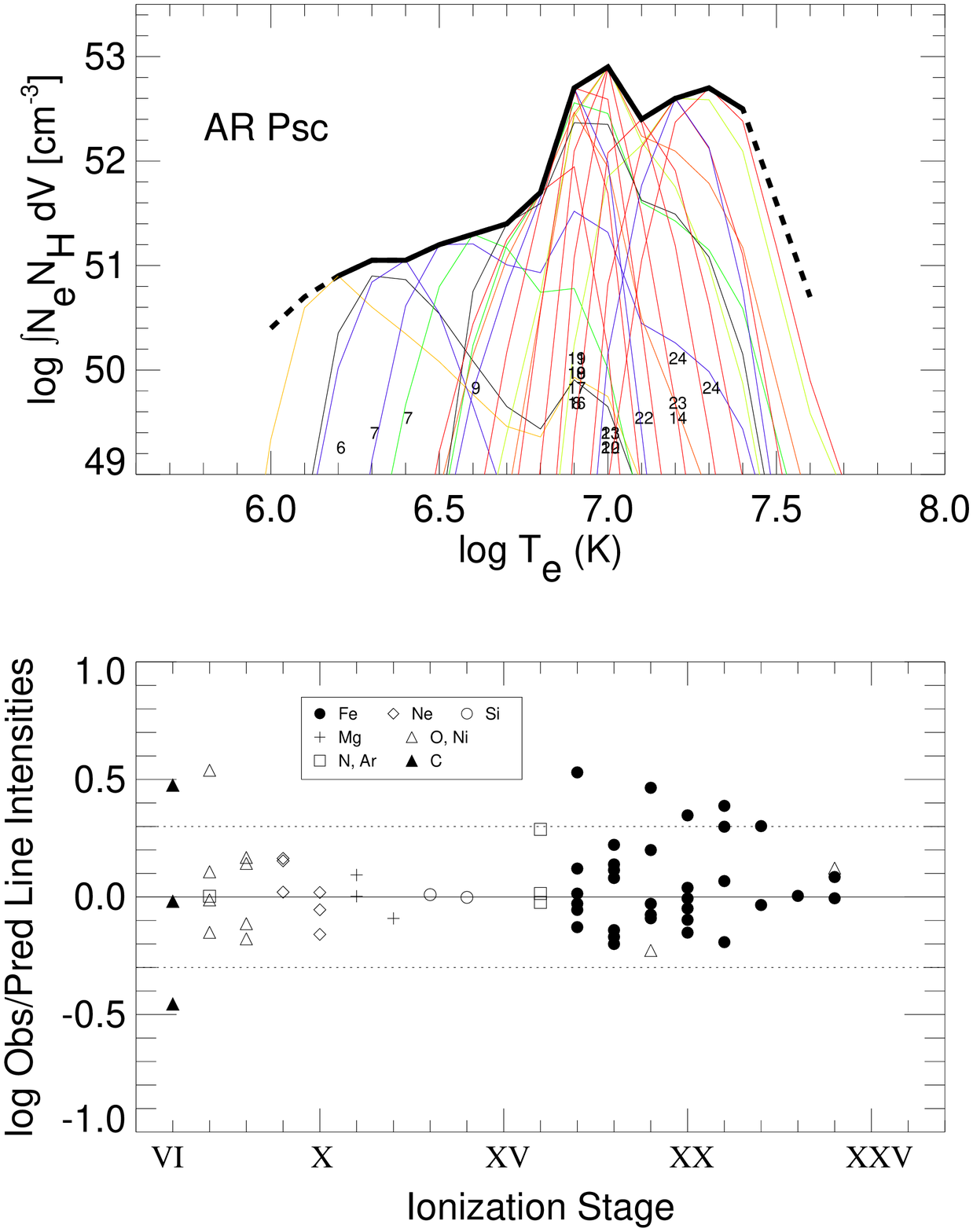}
   \includegraphics[width=0.498\textwidth]{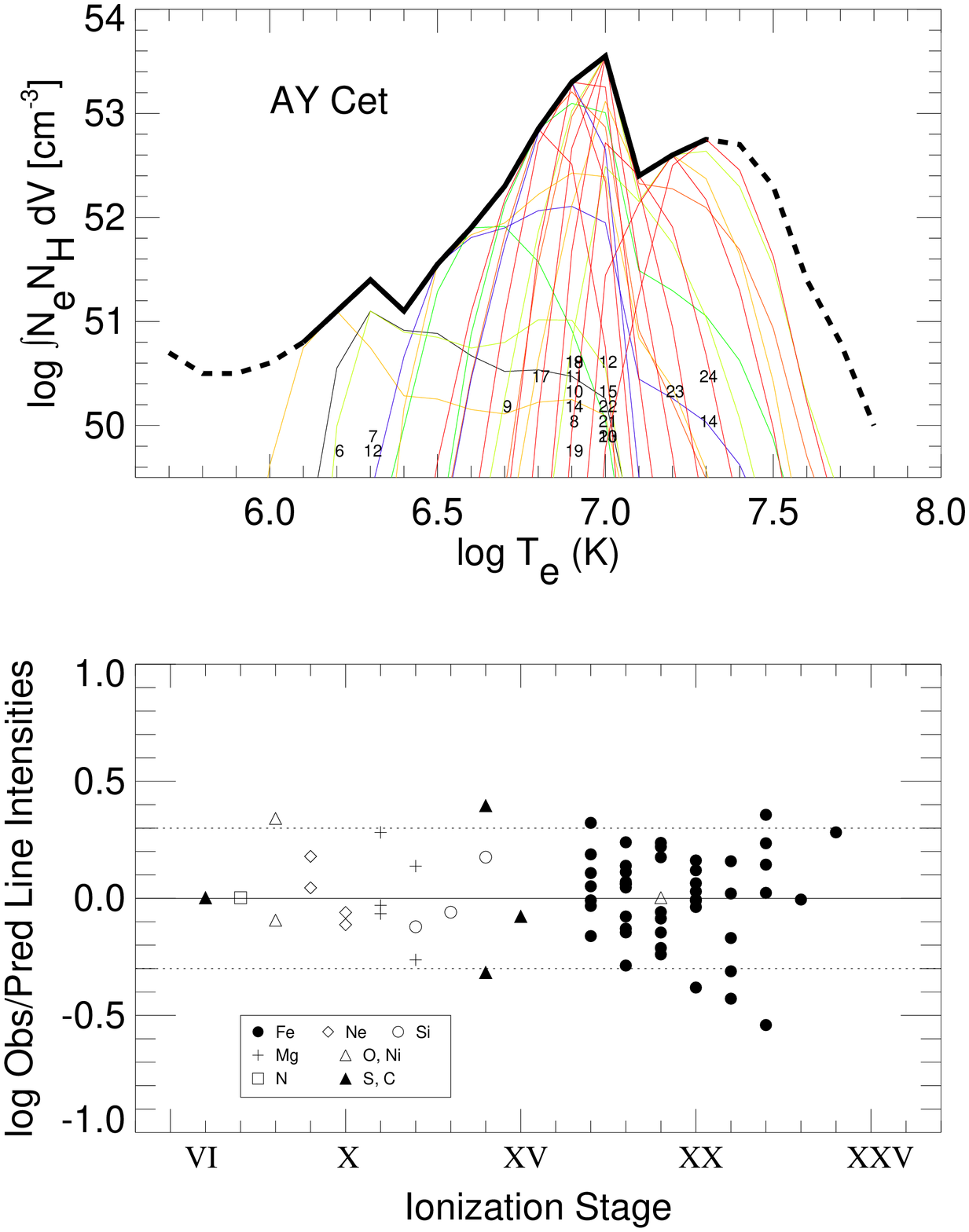}
   \caption{Emission measure distribution (EMD) of AR Psc and AY
     Cet. Thin lines represent the relative
   contribution function for each ion (the emissivity function
   multiplied by the EMD at each point). Small numbers indicate the
   ionization stages of the species. Also plotted are the 
   observed-to-predicted line flux ratios for the ion stages in the 
   upper figure.
   The dotted lines denote a factor of 2.} 
   \label{fig:emds}
\end{figure*}

\input{12069t2.tex}
\input{12069t3.tex}

\begin{table}
\caption{Emission measure distribution of the target stars}\label{tab:emds}
\begin{center}
\begin{small}
\begin{tabular}{lrr}
\hline \hline
{log~$T$} & \multicolumn{2}{c}{log $\int N_{\rm e} N_{\rm H} {\rm d}V$
  (cm$^{-3}$)$^a$} \\
(K) & {AR Psc} & {AY Cet} \\ 
\hline
6.0 & 50.40\hfill\hspace{1ex}  & 50.60\hfill\hspace{1ex}  \\
6.1 & 50.70\hfill\hspace{1ex}  & 50.80\hfill\hspace{1ex}  \\
6.2 & 50.90$^{+0.10}_{-0.30}$  & 51.10$^{+0.20}_{-0.40}$  \\
6.3 & 51.05$^{+0.10}_{-0.30}$  & 51.40$^{+0.20}_{-0.20}$  \\
6.4 & 51.05$^{+0.20}_{-0.30}$  & 51.10$^{+0.30}_{-0.20}$  \\
6.5 & 51.20$^{+0.10}_{-0.30}$  & 51.55$^{+0.30}_{-0.30}$  \\
6.6 & 51.30$^{+0.20}_{-0.30}$  & 51.90$^{+0.30}_{-0.30}$  \\
6.7 & 51.40$^{+0.20}_{-0.30}$  & 52.30$^{+0.20}_{-0.30}$  \\
6.8 & 51.70$^{+0.20}_{-0.20}$  & 52.85$^{+0.20}_{-0.20}$  \\
6.9 & 52.70$^{+0.10}_{-0.00}$  & 53.30$^{+0.10}_{-0.10}$  \\
7.0 & 52.90$^{+0.00}_{-0.10}$  & 53.55$^{+0.00}_{-0.00}$  \\
7.1 & 52.40$^{+0.20}_{-0.20}$  & 52.40$^{+0.15}_{-0.25}$  \\
7.2 & 52.60$^{+0.20}_{-0.30}$  & 52.60$^{+0.10}_{-0.30}$  \\
7.3 & 52.70$^{+0.10}_{-0.30}$  & 52.75$^{+0.15}_{-0.15}$  \\
7.4 & 52.50$^{+0.10}_{-0.30}$  & 52.70$^{+0.30}_{-0.25}$  \\
7.5 & 51.60\hfill\hspace{1ex}  & 52.30\hfill\hspace{1ex} \\
\hline
\end{tabular}
\end{small}
\end{center}
$^a$Emission measure, where $N_{\rm e}$ 
and $N_{\rm H}$ are electron and hydrogen densities, in
cm$^{-3}$. Error bars provided are not independent
between the different temperatures, see text.
\end{table}

\section{Results}
\begin{figure}
   \centering
   \includegraphics[width=0.45\textwidth]{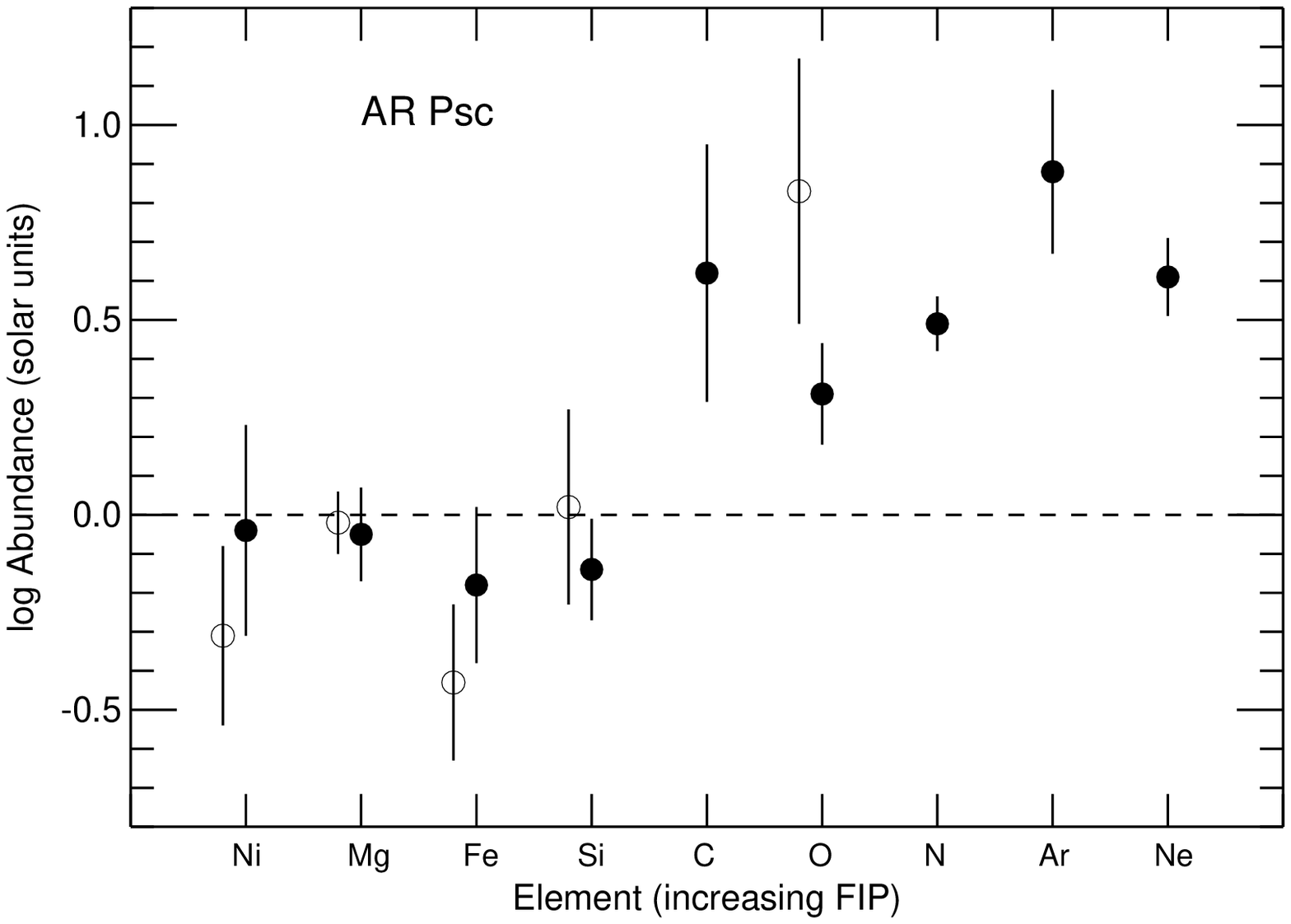}
   \includegraphics[width=0.45\textwidth]{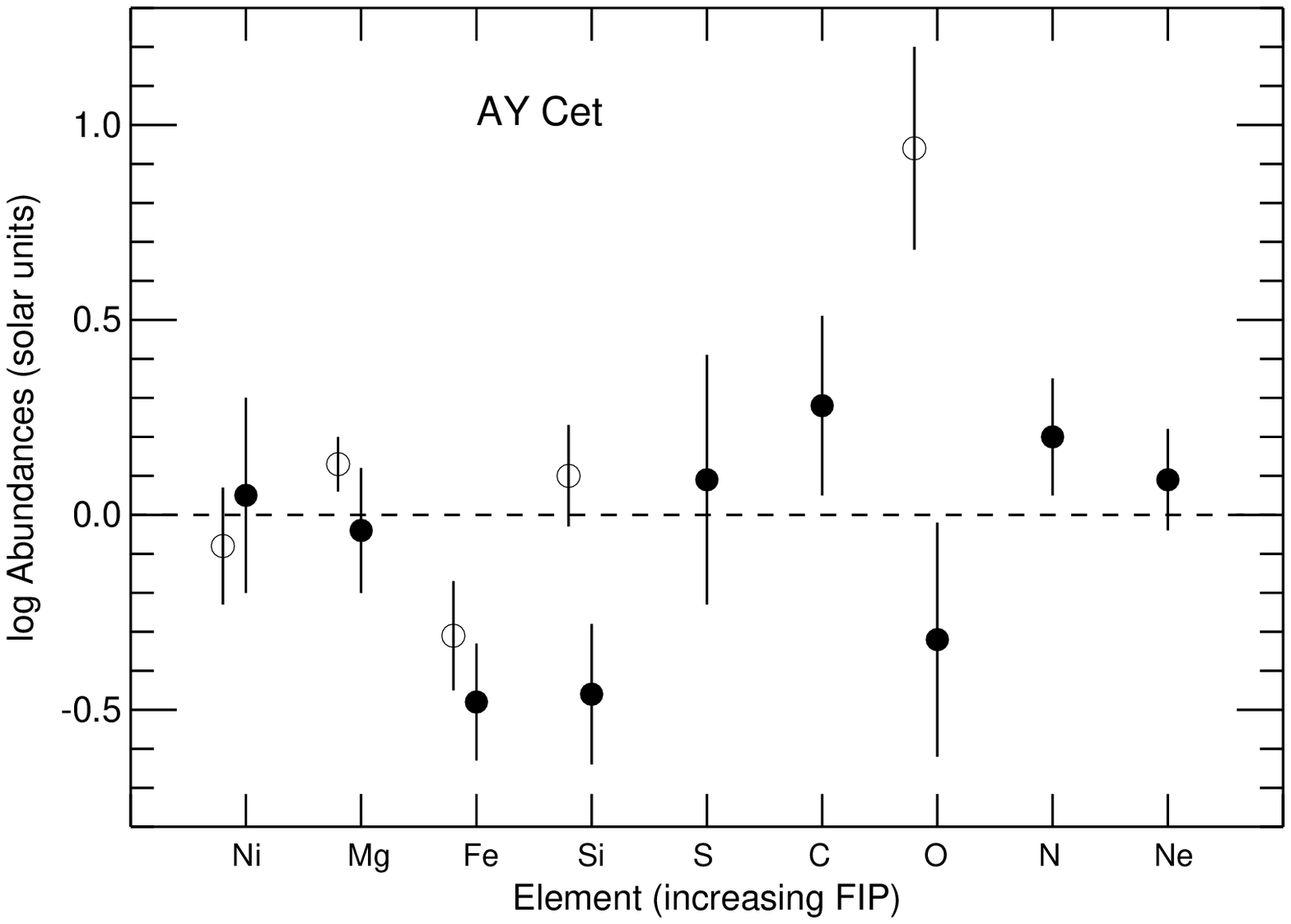}
   \caption{Coronal abundances of AR Psc and AY Cet, in increasing
   order of FIP. Filled circles are coronal values and open circles
   represent photospheric values. A dashed line
   indicates the adopted solar photospheric abundance \citep{asp05}.}  
   \label{fig:abuncorona}
\end{figure}

Light curves of AR Psc and AY Cet (Fig.~\ref{fig:lcs}) show no flares 
in these observations, although variability up to a $\sim$10\% takes place
in AR~Psc. The small portion of the orbital period
covered prevents us from
identifying variability related to orbital phase in both stars. 
Both systems behave like the EUVE observations reported by 
\citet{sanz03}. The spectra recorded by EUVE were of limited use because of
low statistics. The large number of lines observed with XMM-Newton
and Chandra have allowed us to construct a more accurate emission
measure distribution (EMD) as a
function of temperature, defined as $E\!M(T) = \int_{\Delta T} N_H N_e
dV$ [cm$^{-3}$]. We used a
line-based analysis described in \citet{sanz03b} and references
therein. In short,
individual line fluxes are measured\footnote{We then measured fluxes
  for interstellar medium absorption (ISM), using $\log N_{\rm
    H}=18.8$ (AY Cet) and $\log N_{\rm H}=18.3$ (AR Psc), although it
  is only important in a few lines.} 
(Tables~\ref{tab:flarpsc}, \ref{tab:flaycet}) and then compared to a
trial EMD, which is combined with the atomic emission model
Astrophysical Plasma Emission Database \citep[APED v1.3.1,][]{aped} in
order to produce theoretical fluxes of the lines. The comparison of
measured and modeled line fluxes result in an improved EMD that is
used again to produce new modeled line fluxes. 
The iterative process result
in a solution that is not unique, but reliably approximates the
observed and modeled fluxes, and it presumably resembles the real EMD
of the corona. Error bars were calculated using a Monte Carlo method
that seeks the best solution for different line fluxes within their
1--$\sigma$ errors \citep[see][for more details]{sanz03b}. 
In our case the lines measured are formed
at different temperatures and correspond to several elements. We took
the caution to construct an initial
EMD using only Fe lines and then progressively added lines of each
element with overlapping temperatures to the analysis. The determined
EMDs are displayed in Fig.~\ref{fig:emds} and Table~\ref{tab:emds}.
Coronal abundances (Fig.~\ref{fig:abuncorona}, Table~\ref{tab:abundances}) 
were calculated with the same process and then
compared to the values measured in their own
photospheres (see below). We used the solar photospheric values \citep{asp05} 
as reference. Some solar abundance values have
changed between \citet{anders} \citep[used in][]{sanz04} and
\citet{asp05}. Most notably, [Fe/H] is now 7.45 instead of 7.67. 

\begin{figure}
   \centering
   \includegraphics[width=0.45\textwidth]{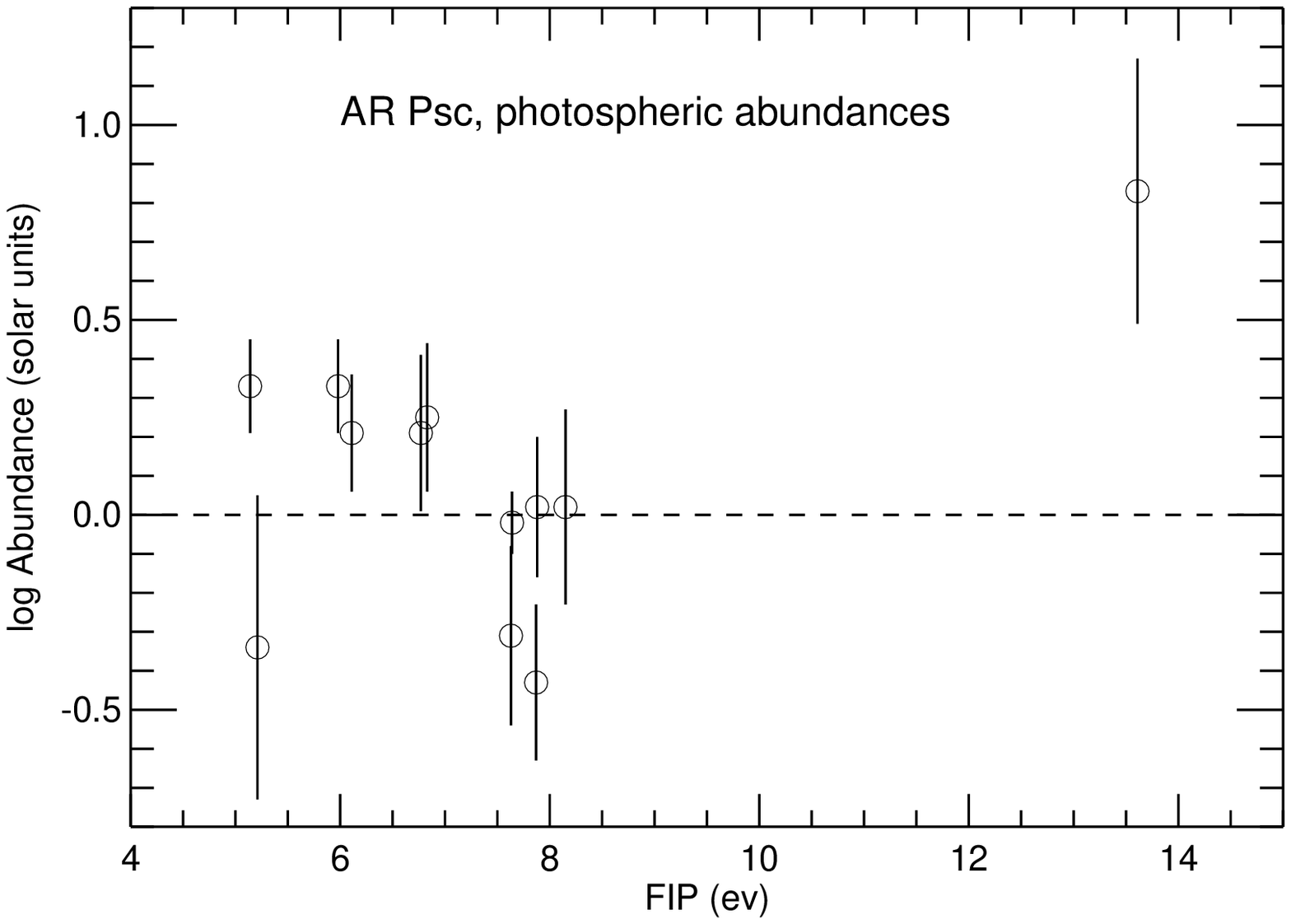}
   \includegraphics[width=0.45\textwidth]{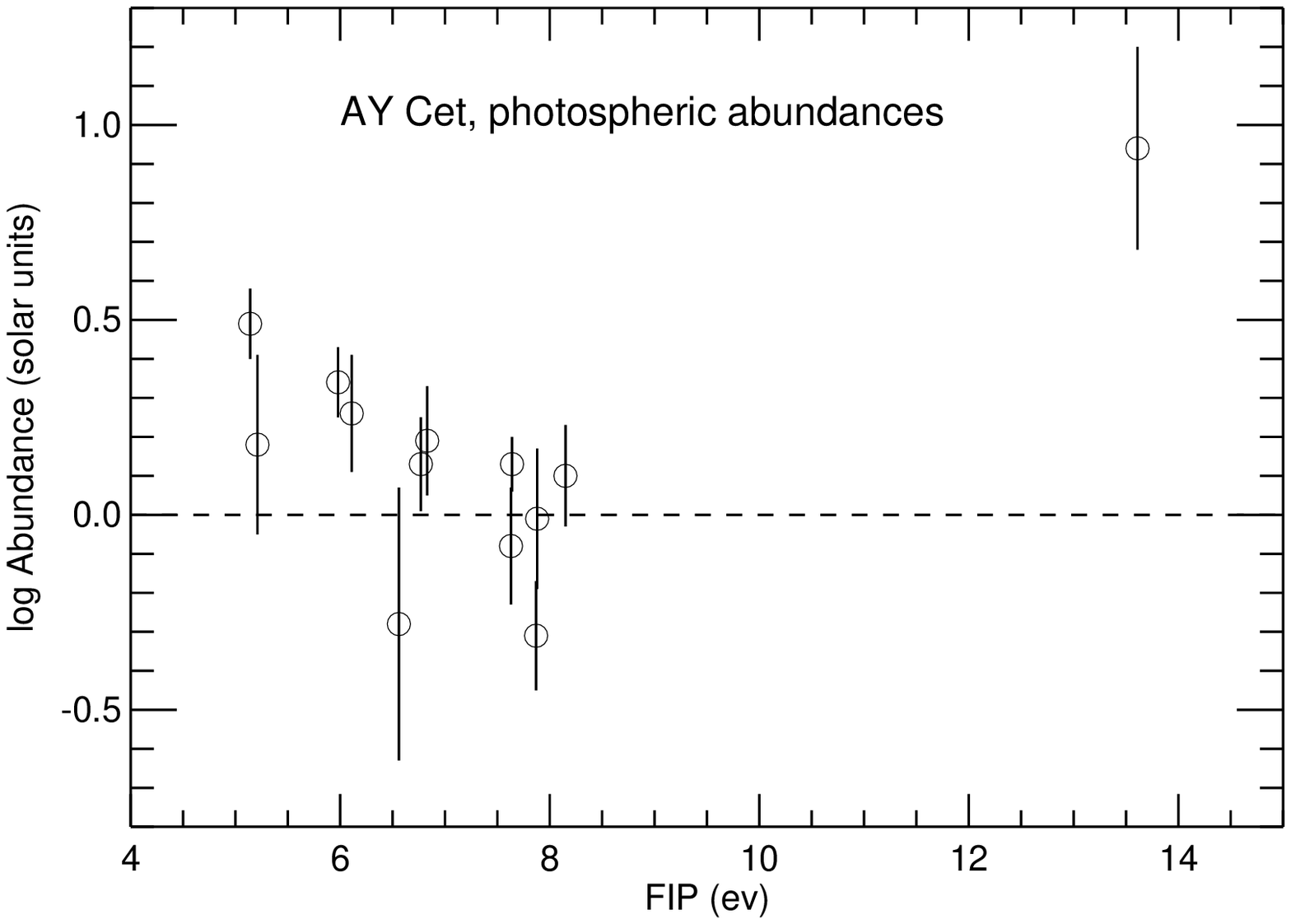}
   \caption{Photospheric abundances of AR Psc and AY Cet. A dashed line
   indicates the adopted solar photospheric abundance \citep{asp05}.} 
   \label{fig:abunphot}
\end{figure}

The photospheric abundances (Fig.~\ref{fig:abunphot}, 
Table~\ref{tab:abundances}) were calculated
through the analysis of the 
optical spectra, in which we select a group of lines that are free
of blends, as explained in \citet{aff05} and \citet{mor03}.
To avoid the difficulty in defining the continuum in
the blue part of the spectra, only lines with $\lambda$ $>$ 5500 \AA\ were
selected. Lines that appeared
asymmetric or showed an unusually large width were assumed to be blended with
unidentified lines so were discarded from the initial sample.
To obtain information on individual abundances from the spectral
lines of various elements, one must first 
determine the parameters that characterize the atmospheric model; i.e., the
effective temperature ($T_{\rm eff}$), the surface gravity ($\log g$),
the microturbulent velocity ($\xi$), and the
iron abundance. They were calculated in an iterative process from the
comparison of the observed spectra and the model for a given set of
parameters. 
The atmospheric parameters and metal
abundances were determined using the measured $EW$s and a standard
local thermodynamic equilibrium (LTE) analysis with the most recent
version of the line abundance code 
MOOG \citep{sne73} and a grid of \citet{kur93} ATLAS9 atmospheres,
computed without the overshooting option and with a mixing length to
pressure scale height ratio $\alpha 
=0.5$. 
Assumptions made in the models include: the
atmosphere is plane-parallel and in hydrostatic equilibrium, the
total flux is constant, the source function
is described by the Planck function, the populations of different excitation
levels and ionization stages are governed by LTE.
The abundances were derived from theoretical curves of growth, computed
by MOOG, using model atmospheres 
and atomic data (wavelength, 
excitation potential, $gf$ values). The input model was constructed using as
atmospheric parameters the average values of previous determinations
found in the literature and solar metallicity.
Further details on the iterative process followed, and the errors
determination is described in \citet{aff05}.
For an elemental abundance derived
from many lines, the uncertainty of the atmospheric parameters is the dominant
error, while for an abundance derived from a few lines, the uncertainty in the
equivalent widths may be more significant.
In our case, the atmospheric parameters calculated are 
$T_{\rm eff}=4995\pm 170$~K, $\log g=3.27\pm 0.60$ [cm\,s$^{-2}$] ,
and $\xi=1.29\pm 0.13$~km\,s$^{-1}$  for AR Psc, and $T_{\rm eff}
=4967\pm 185$~K, $\log g=2.34 \pm 0.47$ [cm\,s$^{-2}$], and $\xi=1.56
\pm 0.08$~km\,s$^{-1}$ 
for AY Cet. In the case of AY Cet, these parameters are in good
agreement with \citet{ott98} except for lower gravity in our
case; iron abundance agrees with that of Ottman et al., but Mg and Si
are substantially higher (-0.22 and -0.32 in their case, respectively).

\begin{figure*}
   \centering
   \includegraphics[width=0.45\textwidth]{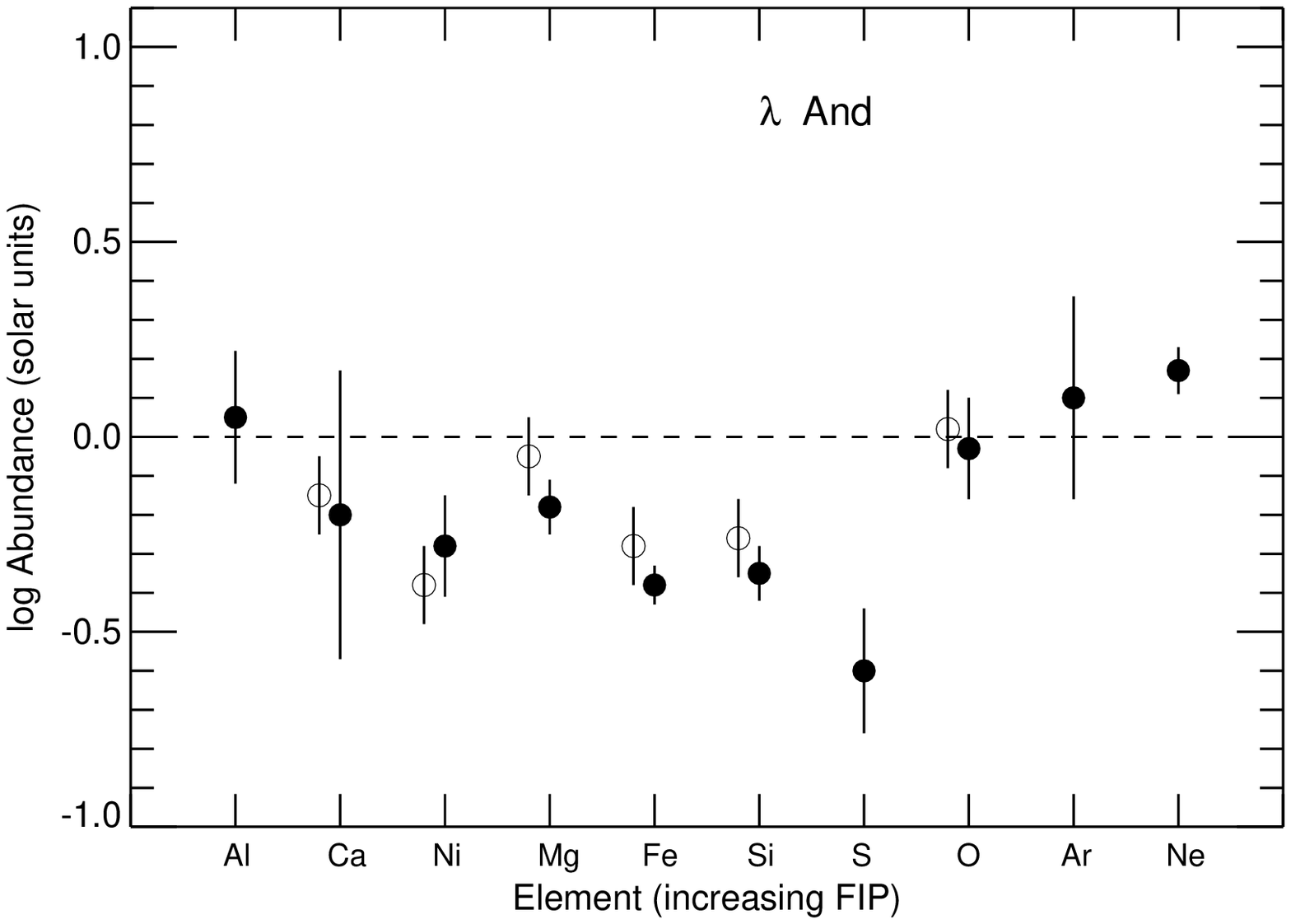}
   \includegraphics[width=0.45\textwidth]{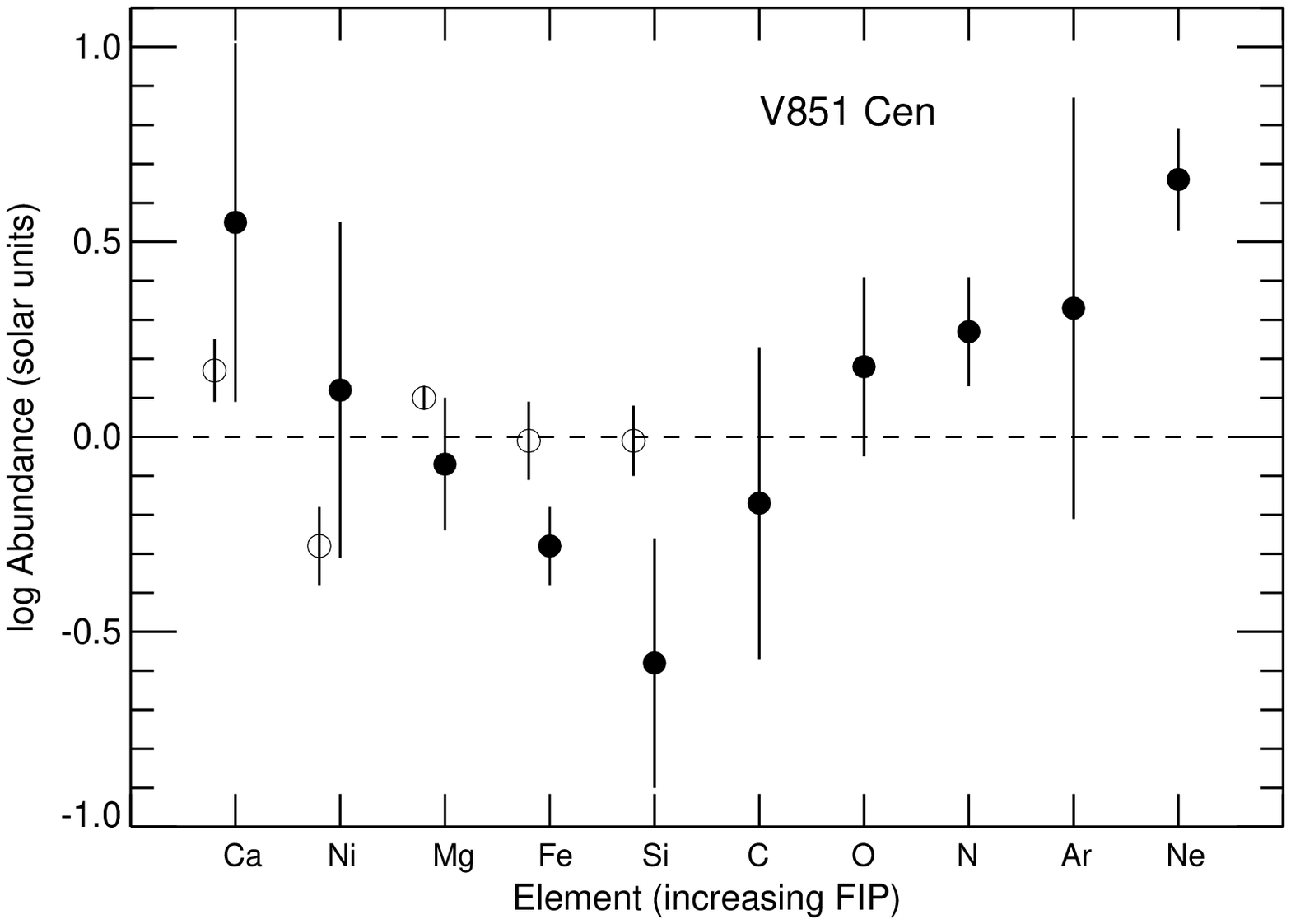}
   \caption{Same as in Fig.~\ref{fig:abuncorona}, but for
     $\lambda$~And and V851 Cen. Results from \citet{sanz04}, adapted to
     the solar reference values of \citet{asp05}.}
   \label{fig:abuncorona2}
\end{figure*}

The coronal abundances are compared to the photosphere of the same
stars (Fig.~\ref{fig:abuncorona}), using the solar photospheric values
as scale. There is no proof of 
MAD or any inverse FIP effect in either of the two cases. 
The best calculated values, the abundance of Fe, Ni and Mg,
show consistent coronal and photospheric abundances. 
Some metal depletion seems to take place for Si and O in AY Cet.
It must be noticed that photospheric
abundances of elements other than Fe are calculated with fewer lines, so
we are less confident about their values.
These results confirm those of $\lambda$~And and V851~Cen
\citep{sanz04}, displayed here for easier comparison
(Fig.~\ref{fig:abuncorona2}, Table~\ref{tab:abundances}.
It is also worth checking the abundance ratio [Ne/O] in the
  corona. \citet{dra05} find that an average value of [Ne/O]=0.41,
  calculated in the corona of nearby stars, would help for solving the
  ``solar model problem'' \citep[see also discussion in][]{sch05}. The
  ratios in the coronae of AR~Psc and AY~Cet are consistent with the
  value observed by \citet{dra05}. An upwards correction of the Ne
  solar abundance, as proposed by these authors, would not affect our
  results since we compared coronal and photospheric abundances of the
  same star.

\section{Discussion}

The results are suggestive of a lack of MAD effect. 
We see once again that coronal
abundances of active stars, once they are compared with their own
photospheric values instead of the solar photosphere, show no sign of
inverse FIP effect or MAD. So far, all the active stars with inverse FIP
effect observed (we discard those compared to solar photosphere) have
a high projected rotational velocity 
(Table~\ref{tab:pastabundances}). A large $v \sin i$ broadens the
lines observed in the optical wavelengths, those usually employed to 
calculate photospheric abundances. 
The broadening of lines might yield erroneous measurements of equivalent
widths (blends are included in the measurements, and continuum is more
difficult to place), and
therefore wrong photospheric abundances. 
All active stars are fast rotators, but only the cases with high projected
rotational velocity will 
affect the measurements of lines. \citet{ott98} note that values of
$v \sin i \ga 20$\,km\,s$^{-1}$ yield wrong results. We therefore are more inclined
to believe that the inverse FIP effect is an observational effect,
rather than a real fact. Thus, from now on we will interpret the
results in this basis, insisting that more observations are needed in order to
clarify how real the inverse FIP effect is.

\begin{table*}
\caption{Coronal and photospheric abundances of the elements ([X/H], solar
  units) in the target stars.}\label{tab:abundances} 
\tabcolsep 3.pt
\begin{center}
\begin{footnotesize}
 \begin{tabular}{lrccrrrrrrrr}
\hline \hline
{X} & {FIP} & Ref.$^a$ & (AG89$^a$) & \multicolumn{2}{c}{AR Psc} &
\multicolumn{2}{c}{AY Cet} & \multicolumn{2}{c}{$\lambda$ And$^b$} &
\multicolumn{2}{c}{V851 Cen$^b$}\\ 
    &  eV   & \multicolumn{2}{c}{solar photosphere} & Photosphere & Corona & Photosphere & Corona & Photosphere & Corona & Photosphere & Corona\\
\hline
 Na &  5.14 &  6.17 & (6.33) &  0.33$\pm$ 0.12 & \ldots &  0.49$\pm$ 0.09 & \ldots & -0.09$\pm$ 0.10 & \ldots &  0.39$\pm$ 0.11 & \ldots \\
 Al &  5.98 &  6.37 & (6.47) &  0.33$\pm$ 0.12 & \ldots &  0.34$\pm$ 0.09 & \ldots & \ldots &  0.05$\pm$ 0.17 &  0.35$\pm$ 0.05 & \ldots \\
 Ca &  6.11 &  6.31 & (6.36) &  0.21$\pm$ 0.15 & \ldots &  0.26$\pm$ 0.15 & \ldots & -0.15$\pm$ 0.10 & -0.20$\pm$ 0.37 &  0.17$\pm$ 0.08 &  0.55$\pm$ 0.46 \\
 Ni &  7.63 &  6.23 & (6.25) & -0.31$\pm$ 0.23 & -0.04$\pm$ 0.27 & -0.08$\pm$ 0.15 &  0.05$\pm$ 0.25 & -0.38$\pm$ 0.10 & -0.28$\pm$ 0.13 & -0.28$\pm$ 0.10 &  0.12$\pm$ 0.43 \\
 Mg &  7.64 &  7.53 & (7.58) & -0.02$\pm$ 0.08 & -0.05$\pm$ 0.12 &  0.13$\pm$ 0.07 & -0.04$\pm$ 0.16 & -0.05$\pm$ 0.10 & -0.18$\pm$ 0.07 &  0.10$\pm$ 0.03 & -0.07$\pm$ 0.17 \\
 Fe &  7.87 &  7.45 & (7.67) & -0.43$\pm$ 0.20 & -0.18$\pm$ 0.20 & -0.31$\pm$ 0.14 & -0.48$\pm$ 0.15 & -0.28$\pm$ 0.10 & -0.38$\pm$ 0.05 & -0.01$\pm$ 0.10 & -0.28$\pm$ 0.10 \\
 Si &  8.15 &  7.51 & (7.55) &  0.02$\pm$ 0.25 & -0.14$\pm$ 0.13 &  0.10$\pm$ 0.13 & -0.46$\pm$ 0.18 & -0.26$\pm$ 0.10 & -0.35$\pm$ 0.07 & -0.01$\pm$ 0.09 & -0.58$\pm$ 0.32 \\
  S & 10.36 &  7.14 & (7.21) & \ldots & \ldots & \ldots &  0.09$\pm$ 0.32 & \ldots & -0.60$\pm$ 0.16 & \ldots & -0.98$\pm$ 1.32 \\
  C & 11.26 &  8.39 & (8.56) & \ldots &  0.62$\pm$ 0.33 & \ldots &  0.28$\pm$ 0.23 & \ldots & \ldots & \ldots & -0.17$\pm$ 0.40 \\
  O & 13.61 &  8.66 & (8.93) &  0.83$\pm$ 0.34 &  0.31$\pm$ 0.13 &  0.94$\pm$ 0.26 & -0.32$\pm$ 0.30 &  0.02$\pm$ 0.10 & -0.03$\pm$ 0.13 & \ldots &  0.18$\pm$ 0.23 \\
  N & 14.53 &  7.78 & (8.05) & \ldots &  0.49$\pm$ 0.07 & \ldots &  0.20$\pm$ 0.15 & \ldots & \ldots & \ldots &  0.27$\pm$ 0.14 \\
 Ar & 15.76 &  6.18 & (6.56) & \ldots &  0.88$\pm$ 0.21 & \ldots & \ldots & \ldots &  0.10$\pm$ 0.26 & \ldots &  0.33$\pm$ 0.54 \\
 Ne & 21.56 &  7.84 & (8.09) & \ldots &  0.61$\pm$ 0.10 & \ldots &  0.09$\pm$ 0.13 & \ldots &  0.17$\pm$ 0.06 & \ldots &  0.66$\pm$ 0.13 \\
\hline
\end{tabular}
\end{footnotesize}
\end{center}
$^a$ Solar photospheric abundances from \citet{asp05}, adopted in
this work, are expressed in logarithmic scale. 
Note that several values have been
updated in the literature since \citet{anders}, also listed in
parenthesis for easier comparison.\\
$^b$ Results adapted from \citet{sanz04}.
\end{table*}

Assuming that no inverse FIP effect exists among active stars, the
observations suits the coronal model explained by
\citet{lam04}. The works carried out in the Sun in past years have
strengthened the idea of the Alfv\'en waves as the main force responsible for
the energy transportation between chromosphere and corona of the
Sun. Recent observations with the mission Hinode \citep{dep07,erd07}
support this idea. In the model described by \citet{lam04}, the
Alfv\'en waves, combined with 
pondermotive forces, would be responsible for
part of the energy transportation in the corona. Their model explains
the observed abundances pattern in the solar corona: either the
low FIP in the coronal active regions and the slow wind or the
absence of FIP effect in coronal holes or fast wind. According to this
model, the stars earlier than the Sun, like Procyon, would behave more
like coronal holes, because they have shallower convective zones
yielding smaller coronal magnetic fields. Their chromospheric Alfv\'en
waves are stronger than the coronal fields, similar to a 
coronal hole. In stars later than the Sun, but not very active yet,
there is a deeper convection, yielding lower frequency chromospheric
waves and higher coronal magnetic fields. The wave transmission is less
efficient and this would yield a reduced FIP effect. This is the case
for $\epsilon$~Eri or $\alpha$~Cen.
Following the same reasoning, it makes
sense to find that Alfv\'en waves are more inefficient for the more
active stars, yielding no FIP fractionation for stars such as AR~Psc
or AY~Cet. Some arguments have been
proposed by \citet{lam04} to explain why an inverse FIP effect could
be present in active stars, such as AB~Dor: wave reflection in the
chromosphere (more 
likely for stars with lower gravity), or turbulence introduced by
differential rotation. 

A more complete census of active stars with well measured photospheric
and coronal abundances would be necessary to confirm the results, but
if we assume that the cases with inverse FIP effect observed are a
product of observational effects, the model proposed by \citet{lam04}
could actually explain the observed abundances patterns in both low
and high activity stars.

\section{Conclusions}

The results suggest that metal abundance depletion, or
inverse FIP effect, are not present in these two stars. This
agrees with results for other active stars with
well-known photospheric abundances, such as $\lambda$~And or 
V851~Cen. Stars with observed inverse FIP effect have their
photospheric lines broadened by the high rotation of the stars,
hampering their photospheric measurements. The results observed in
AR~Psc and AY~Cet suit the expected behavior in active stars
according to the model of \citet{lam04}, which combines the use of
pondermotive forces with Alfv\'en waves. The same model is also able to
explain both the FIP effect in the solar corona and slow wind and the
absence of FIP fractionation in solar coronal holes and fast wind.
Further measurements in active
stars with low projected rotational velocity would help to assure this
confirmation of the model. 

\begin{acknowledgements}
  We acknowledge support by the Ram\'on y Cajal Program
  ref. RYC-2005-000549, financed by the Spanish Ministry of Science.
  JS wants to thank Aitor Ibarra for his aid in the treatment of XMM
  data. This research has also made use
  of NASA's Astrophysics Data System Abstract Service and the SIMBAD
  Database, operated at CDS, Strasbourg (France).
\end{acknowledgements}



\Online

\end{document}

%% file: 12069t2.tex
\onltab{2}{
\begin{table*}
\caption{XMM/RGS line fluxes of AR Psc$^a$}\label{tab:flarpsc}
\tabcolsep 3.pt
\begin{scriptsize}
\begin{tabular}{lrcrrrl}
\hline \hline
 Ion & {$\lambda$$_{\mathrm {model}}$} &  
 log $T_{\mathrm {max}}$ & $F_{\mathrm {obs}}$ & S/N & ratio & Blends \\ 
\hline
\ion{Si}{xiv} &  6.1804 & 7.2 & 8.06e-14 &   4.3 & -0.00 &  \\
\ion{Si}{xiii} &  6.6479 & 7.0 & 2.43e-13 &  10.2 &  0.01 & \ion{Si}{xiii}  6.6882, 6.7403 \\
\ion{Mg}{xii} &  8.4192 & 7.0 & 1.59e-13 &   5.9 & -0.09 & \ion{Mg}{xii}  8.4246 \\
\ion{Mg}{xi} &  9.1687 & 6.8 & 1.12e-13 &  10.8 &  0.00 & \ion{Mg}{xi}  9.2312 \\
\ion{Mg}{xi} &  9.3143 & 6.8 & 5.11e-14 &   7.5 &  0.09 & \ion{Ni}{xxv}  9.3400, \ion{Fe}{xxii}  9.3933 \\
\ion{Fe}{xxi} &  9.4797 & 7.0 & 1.15e-13 &  11.5 &  0.30 & \ion{Ne}{x}  9.4807,  9.4809, \ion{Ni}{xxvi}  9.5292, \ion{Fe}{xxi}  9.5443 \\
\ion{Ne}{x} &  9.7080 & 6.8 & 4.34e-14 &   7.2 & -0.16 & \ion{Ne}{x} 9.7085, \ion{Fe}{xx}  9.7269 \\
\ion{Fe}{xx} & 10.1203 & 7.0 & 5.74e-14 &   8.6 &  0.35 & \ion{Fe}{xx} 10.0529, 10.0596, \ion{Ni}{xix} 10.1100, \ion{Fe}{xix} 10.1419 \\
\ion{Ne}{x} & 10.2385 & 6.8 & 1.55e-13 &   8.2 &  0.02 & \ion{Ne}{x} 10.2396 \\
\ion{Ni}{xxiv} & 10.2770 & 7.2 & 1.23e-14 &   4.1 &  0.12 & \ion{Fe}{xx} 10.2675 \\
\ion{Fe}{xxiv} & 10.6190 & 7.3 & 8.69e-14 &  11.3 & -0.01 & \ion{Fe}{xix} 10.6414, 10.6491, \ion{Fe}{xxiv} 10.6630 \\
\ion{Fe}{xix} & 10.8160 & 6.9 & 2.33e-14 &   5.9 & -0.08 & \ion{Ne}{ix} 10.7650, \ion{Fe}{xvii} 10.7700 \\
\ion{Fe}{xxiii} & 10.9810 & 7.2 & 1.11e-13 &  13.1 &  0.00 & \ion{Ne}{ix} 11.0010, \ion{Fe}{xxiii} 11.0190, \ion{Fe}{xxiv} 11.0290 \\
\ion{Fe}{xxiv} & 11.1760 & 7.3 & 7.44e-14 &  10.9 &  0.09 & \ion{Fe}{xvii} 11.1310, \ion{Fe}{xxiv} 11.1870 \\
\ion{Fe}{xvii} & 11.2540 & 6.8 & 2.22e-14 &   6.0 & -0.05 & \ion{Fe}{xxiv} 11.2680, \ion{Fe}{xxiii} 11.2850 \\
\ion{Fe}{xviii} & 11.4230 & 6.9 & 4.36e-14 &   3.4 & -0.17 & \ion{Fe}{xxii} 11.4270, \ion{Fe}{xxiv} 11.4320, \ion{Fe}{xxiii} 11.4580 \\
\ion{Ne}{ix} & 11.5440 & 6.6 & 9.44e-14 &  12.1 &  0.15 & \ion{Fe}{xxii} 11.4900, \ion{Fe}{xviii} 11.5270, \ion{Ni}{xix} 11.5390 \\
\ion{Fe}{xxii} & 11.7700 & 7.1 & 1.86e-13 &   9.4 & -0.03 & \ion{Fe}{xxiii} 11.7360 \\
\ion{Fe}{xxii} & 11.9770 & 7.1 & 1.25e-13 &   5.7 &  0.30 & \ion{Fe}{xxii} 11.8810,  11.9320, \ion{Fe}{xxiii} 11.8980, \ion{Fe}{xxi} 11.9466, 11.9750 \\
\ion{Ne}{x} & 12.1321 & 6.8 & 8.36e-13 &  19.6 & -0.05 & \ion{Ne}{x} 12.1375 \\
\ion{Fe}{xxi} & 12.2840 & 7.0 & 1.31e-13 &   7.7 & -0.19 & \ion{Fe}{xvii} 12.2660 \\
\ion{Ni}{xix} & 12.4350 & 6.9 & 5.83e-14 &  10.6 & -0.23 & \ion{Fe}{xxi} 12.3930, 12.4220, \ion{Fe}{xxii} 12.4311, 12.4318 \\
\ion{Fe}{xxi} & 12.4990 & 7.0 & 5.01e-14 &   3.6 &  0.39 & \ion{Ni}{xxi} 12.5105, \ion{Fe}{xx} 12.5260 \\
\ion{Fe}{xx} & 12.8460 & 7.0 & 1.73e-13 &  10.9 & -0.15 & \ion{Fe}{xxi} 12.8220, \ion{Fe}{xx} 12.8240, 12.8640 \\
\ion{Fe}{xx} & 12.9650 & 7.0 & 8.99e-14 &  12.9 & -0.10 & \ion{Fe}{xx} 12.9120, 12.9920, 13.0240, \ion{Fe}{xix} 12.9330, 13.0220, \ion{Fe}{xxii} 12.9530 \\
\ion{Fe}{xx} & 13.1530 & 7.0 & 4.12e-14 &   8.5 & -0.05 & \ion{Fe}{xx} 13.1370, \ion{Fe}{xxi} 13.1155, 13.1671, 13.1678 \\
\ion{Fe}{xx} & 13.2740 & 7.0 & 3.75e-14 &   3.1 & -0.01 & \ion{Fe}{xx} 13.2909, \ion{Fe}{xxii} 13.2360, \ion{Fe}{xxi} 13.2487, \ion{Ni}{xx} 13.2560, \ion{Fe}{xix} 13.2658 \\
\ion{Fe}{xx} & 13.3089 & 7.0 & 1.88e-14 &   6.2 &  0.04 & \ion{Ni}{xx} 13.3090, \ion{Fe}{xix} 13.3191, \ion{Fe}{xviii} 13.3230 \\
\ion{Ne}{ix} & 13.4473 & 6.6 & 2.47e-13 &   9.6 &  0.02 & \ion{Fe}{xx} 13.3850, \ion{Fe}{xix} 13.4620 \\
\ion{Fe}{xix} & 13.5180 & 6.9 & 1.74e-13 &  19.2 & -0.09 & \ion{Fe}{xix} 13.4970, \ion{Fe}{xxi} 13.5070, \ion{Ne}{ix} 13.5531 \\
\ion{Ne}{ix} & 13.6990 & 6.6 & 1.69e-13 &  19.0 &  0.16 & \ion{Fe}{xix} 13.6450, 13.7315, 13.7458 \\
\ion{Fe}{xix} & 13.7950 & 6.9 & 7.97e-14 &  13.1 & -0.03 & \ion{Fe}{xx} 13.7670, \ion{Ni}{xix} 13.7790, \ion{Fe}{xvii} 13.8250 \\
\ion{Fe}{xxi} & 14.0080 & 7.0 & 9.01e-14 &  13.9 &  0.07 & \ion{Ni}{xix} 14.0430,  14.0770, \ion{Fe}{xix} 14.0717 \\
\ion{Fe}{xviii} & 14.2080 & 6.9 & 1.31e-13 &  16.8 & -0.14 & \ion{Fe}{xviii} 14.2560, \ion{Fe}{xx} 14.2670 \\
\ion{Fe}{xviii} & 14.3730 & 6.9 & 1.19e-13 &   6.7 &  0.08 & \ion{Fe}{xx} 14.3318,  14.4207,  14.4600, \ion{Fe}{xviii} 14.3430, 14.4250 \\
\ion{Fe}{xviii} & 14.5340 & 6.9 & 6.86e-14 &   4.8 &  0.14 & \ion{Fe}{xviii} 14.4856, 14.5056,  14.5710, 14.6011, \ion{Fe}{xx} 14.5146 \\
\ion{Fe}{xix} & 14.6640 & 6.9 & 3.92e-14 &   9.1 &  0.20 & \ion{Fe}{xviii} 14.6884 \\
\ion{O}{viii} & 14.8205 & 6.5 & 9.55e-14 &  14.4 &  0.17 & \ion{Fe}{xix} 14.7250, \ion{Fe}{xx} 14.7540,  14.8526, \ion{Fe}{xviii} 14.7820, \ion{O}{viii} 14.8207 \\
\ion{Fe}{xvii} & 15.0140 & 6.7 & 1.85e-13 &  12.3 & -0.13 &  \\
\ion{O}{viii} & 15.1760 & 6.5 & 1.46e-13 &  10.1 &  0.14 & \ion{Fe}{xix} 15.0790,  15.1980, \ion{O}{viii} 15.1765 \\
\ion{Fe}{xvii} & 15.2610 & 6.7 & 7.26e-14 &   6.4 &  0.12 &  \\
\ion{Fe}{xvii} & 15.4530 & 6.7 & 5.98e-14 &   6.2 &  0.53 & \ion{Fe}{xx} 15.4077, 15.5170, \ion{Fe}{xix} 15.4136, 15.4655, \ion{Fe}{xviii} 15.4940, 15.5199  \\
\ion{Fe}{xviii} & 15.8240 & 6.8 & 4.61e-14 &   3.7 &  0.22 & \ion{Fe}{xviii} 15.8700 \\
\ion{O}{viii} & 16.0055 & 6.5 & 1.36e-13 &  10.0 & -0.18 & \ion{Fe}{xviii} 16.0040, \ion{O}{viii} 16.0067 \\
\ion{Fe}{xviii} & 16.0710 & 6.8 & 5.72e-14 &   4.9 & -0.20 & \ion{Fe}{xix} 16.1100, \ion{Fe}{xviii} 16.1590 \\
\ion{Fe}{xix} & 16.2830 & 6.9 & 5.28e-14 &   6.4 &  0.46 & \ion{Fe}{xviii} 16.3200, \ion{Fe}{xix} 16.3414, \ion{Fe}{xvii} 16.3500 \\
\ion{Fe}{xvii} & 16.7800 & 6.7 & 6.91e-14 &  12.6 & -0.03 &  \\
\ion{Fe}{xvii} & 17.0510 & 6.7 & 1.71e-13 &  19.7 &  0.01 & \ion{Fe}{xvii} 17.0960 \\
\ion{Fe}{xviii} & 17.6230 & 6.8 & 3.38e-14 &   4.6 &  0.11 &  \\
\ion{O}{vii} & 17.7680 & 6.4 & 3.95e-14 &   9.3 &  0.54 & \ion{Ar}{xvi} 17.7320, 17.7420 \\
\ion{O}{vii} & 18.6270 & 6.3 & 2.78e-14 &   7.9 & -0.01 & \ion{Ar}{xvi} 18.6240, \ion{Ca}{xviii} 18.6910 \\
\ion{O}{viii} & 18.9671 & 6.5 & 6.91e-13 &  39.9 & -0.11 & \ion{O}{viii} 18.9725 \\
\ion{O}{vii} & 21.6015 & 6.3 & 6.11e-14 &  11.3 & -0.15 &  \\
\ion{O}{vii} & 22.0977 & 6.3 & 6.61e-14 &  11.7 &  0.11 & \ion{Ca}{xvii} 22.1140 \\
\ion{Ar}{xvi} & 23.5460 & 6.7 & 2.92e-14 &   4.3 &  0.01 & \ion{Ar}{xvi} 23.5900 \\
\ion{N}{vii} & 24.7792 & 6.3 & 8.23e-14 &  13.8 &  0.00 & \ion{N}{vii} 24.7846 \\
\ion{Ar}{xvi} & 24.8540 & 6.7 & 2.60e-14 &   3.4 &  0.29 & \\
\ion{Ar}{xvi} & 24.9910 & 6.7 & 2.83e-14 &   7.9 & -0.02 & \ion{Ar}{xvi} 25.0130, \ion{Ar}{xv} 25.0500 \\
\ion{C}{vi} & 26.9896 & 6.2 & 4.11e-14 &   5.7 &  0.48 & \ion{C}{vi} 26.9901 \\
\ion{C}{vi} & 28.4652 & 6.2 & 4.33e-14 &   4.6 & -0.02 & \ion{Ar}{xv} 28.3860, \ion{C}{vi} 28.4663 \\
\ion{C}{vi} & 33.7342 & 6.1 & 7.86e-14 &   6.3 & -0.45 & \ion{C}{vi} 33.7396 \\
\hline
\end{tabular}

{$^a$ Line fluxes (in erg cm$^{-2}$ s$^{-1}$) 
  measured in XMM/RGS AR Psc spectra, and corrected by the ISM absorption. 
  log $T_{\rm max}$ indicates the maximum
  temperature (K) of formation of the line (unweighted by the
  EMD). ``Ratio'' is the log($F_{\mathrm {obs}}$/$F_{\mathrm {pred}}$) 
  of the line. 
  Blends amounting to more than 5\% of the total flux for each line are
  indicated.}
\end{scriptsize}
\end{table*}
}

%% file: 12069t3.tex
\onltab{3}{
\begin{table*}
\caption{Chandra/LETG line fluxes of AY Cet$^a$}\label{tab:flaycet}
\tabcolsep 3.pt
\begin{scriptsize}
\begin{tabular}{lrcrrrl}
\hline \hline
 Ion & {$\lambda$$_{\mathrm {model}}$} &  
 log $T_{\mathrm {max}}$ & $F_{\mathrm {obs}}$ & S/N & ratio & Blends \\ 
\hline
\ion{S}{xv} &   5.1015 & 7.2 & 9.72e-14 &   3.3 & -0.08 & \ion{S}{xv}  5.0387 \\
\ion{Si}{xiv} &   6.1804 & 7.2 & 5.86e-14 &   5.4 &  0.18 & \ion{Si}{xiv}  6.1858 \\
\ion{Si}{xiii} &   6.6479 & 7.0 & 1.11e-13 &   7.9 & -0.06 & \ion{Si}{xiii}  6.6882, 6.7403 \\
\ion{Mg}{xii} &   7.1058 & 7.0 & 4.21e-14 &   4.9 &  0.14 & \ion{Mg}{xii}  7.1069, \ion{Al}{xiii}  7.1710 \\
\ion{Mg}{xi} &   7.8503 & 6.8 & 4.20e-14 &   4.9 &  0.28 & \ion{Al}{xii}  7.8721 \\
\ion{Mg}{xii} &   8.4192 & 7.0 & 9.72e-14 &   7.8 & -0.26 & \ion{Mg}{xii}  8.4246 \\
\ion{Mg}{xi} &   9.1687 & 6.8 & 1.08e-13 &   7.9 & -0.07 &  \\
\ion{Mg}{xi} &   9.3143 & 6.8 & 7.06e-14 &   6.4 & -0.03 & \ion{Mg}{xi}  9.2312, \ion{Ni}{xix}  9.2540 \\
\ion{Fe}{xx} &   9.9977 & 7.0 & 2.48e-14 &   3.8 & -0.04 & \ion{Ni}{xix}  9.9770, \ion{Fe}{xxi}  9.9887, \ion{Fe}{xx} 10.0004, 10.0054 \\
\ion{Ne}{x} &  10.2385 & 6.8 & 4.17e-14 &   4.9 & -0.11 & \ion{Ne}{x} 10.2396 \\
\ion{Fe}{xix} &  10.6414 & 6.9 & 3.10e-14 &   4.2 & -0.09 & \ion{Fe}{xviii} 10.5364, \ion{Fe}{xxiv} 10.6190,  10.6630, \ion{Fe}{xix} 10.6295, 10.6491, 10.6840, \ion{Fe}{xvii} 10.6570 \\
\ion{Fe}{xix} &  10.8160 & 6.9 & 2.07e-14 &   3.5 &  0.18 & \ion{Fe}{xvii} 10.7700 \\
\ion{Fe}{xxiii} &  10.9810 & 7.2 & 3.16e-14 &   4.3 & -0.01 & \ion{Ne}{ix} 11.0010, \ion{Fe}{xxiii} 11.0190, \ion{Fe}{xvii} 11.0260, \ion{Fe}{xxiv} 11.0290 \\
\ion{Fe}{xxiv} &  11.1760 & 7.3 & 7.50e-14 &   6.8 &  0.28 & \ion{Fe}{xvii} 11.1310, \ion{Ni}{xxii} 11.1818, 11.1950, 11.2118 \\
\ion{Fe}{xviii} &  11.3260 & 6.9 & 3.83e-14 &   4.9 &  0.11 & \ion{Fe}{xvii} 11.2540, \ion{Ni}{xxi} 11.3180 \\
\ion{Fe}{xviii} &  11.4230 & 6.9 & 2.61e-14 &   4.1 &  0.07 & \ion{Fe}{xxii} 11.4270, \ion{Fe}{xxiv} 11.4320, \ion{Fe}{xviii} 11.4494 \\
\ion{Fe}{xviii} &  11.5270 & 6.9 & 4.42e-14 &   5.4 &  0.05 & \ion{Fe}{xxii} 11.4900, \ion{Ni}{xix} 11.5390, \ion{Ni}{xxi} 11.5390, \ion{Ne}{ix} 11.5440 \\
\ion{Fe}{xxii} &  11.7700 & 7.1 & 9.48e-14 &   7.9 &  0.02 & \ion{Fe}{xxiii} 11.7360, \ion{Ni}{xx} 11.8320, 11.8460 \\
\ion{Fe}{xxii} &  11.9770 & 7.1 & 8.26e-14 &   7.5 &  0.36 & \ion{Fe}{xxii} 11.8810, 11.9320, \ion{Ni}{xx} 11.9617, \ion{Fe}{xxi} 11.9750, 12.0440 \\
\ion{Ne}{x} &  12.1321 & 6.8 & 3.17e-13 &  14.7 & -0.06 & \ion{Fe}{xvii} 12.1240, \ion{Ne}{x} 12.1375 \\
\ion{Fe}{xxi} &  12.2840 & 7.0 & 1.57e-13 &  10.4 &  0.16 & \ion{Fe}{xxii} 12.2100, \ion{Fe}{xvii} 12.2660 \\
\ion{Fe}{xxi} &  12.3930 & 7.0 & 1.05e-13 &   8.5 &  0.02 & \ion{Fe}{xxi} 12.4220, \ion{Ni}{xix} 12.4350 \\
\ion{Fe}{xx} &  12.5260 & 7.0 & 4.46e-14 &   5.6 &  0.16 & \ion{Fe}{xxi} 12.4990, 12.5698, \ion{Fe}{xx} 12.5760, 12.5760 \\
\ion{Ni}{xix} &  12.6560 & 6.9 & 2.36e-14 &   4.1 &  0.00 & \ion{Fe}{xxi} 12.6490 \\
\ion{Fe}{xxii} &  12.7540 & 7.1 & 2.34e-14 &   4.1 &  0.24 & \ion{Fe}{xvii} 12.6950 \\
\ion{Fe}{xx} &  12.8240 & 7.0 & 1.66e-13 &  10.9 &  0.03 & \ion{Fe}{xxi} 12.8220, \ion{Fe}{xx} 12.8460, 12.8640 \\
\ion{Fe}{xx} &  12.9650 & 7.0 & 8.63e-14 &   7.9 &  0.12 & \ion{Fe}{xx} 12.9120, 12.9920, \ion{Ni}{xx} 12.9270, \ion{Fe}{xix} 12.9330, \ion{Fe}{xxii} 12.9530 \\
\ion{Fe}{xx} &  13.0610 & 7.0 & 4.47e-14 &   5.7 &  0.06 & \ion{Fe}{xix} 13.0220, \ion{Fe}{xx} 13.0240,  13.1000, \ion{Fe}{xxi} 13.0444, 13.0831 \\
\ion{Fe}{xx} &  13.2740 & 7.0 & 3.13e-14 &   4.9 & -0.01 & \ion{Fe}{xxii} 13.2360, \ion{Fe}{xxi} 13.2487, \ion{Ni}{xx} 13.2560, \ion{Fe}{xix} 13.2658, \ion{Fe}{xx} 13.2909 \\
\ion{Fe}{xx} &  13.3850 & 7.0 & 3.90e-14 &   5.5 & -0.01 & \ion{Fe}{xx} 13.3089, 13.3470, \ion{Ni}{xx} 13.3090, \ion{Fe}{xviii} 13.3230, 13.3550, 13.3948 \\
\ion{Ne}{ix} &  13.4473 & 6.6 & 2.61e-13 &  14.2 &  0.05 & \ion{Fe}{xix} 13.4620, 13.4970, 13.5180, \ion{Fe}{xxi} 13.5070 \\
\ion{Fe}{xix} &  13.6450 & 6.9 & 2.31e-14 &   4.3 &  0.24 & \ion{Fe}{xx} 13.6124, 13.6150, \ion{Fe}{xix} 13.6481 \\
\ion{Ne}{ix} &  13.6990 & 6.6 & 5.98e-14 &   6.9 &  0.18 & \ion{Fe}{xix} 13.6742, 13.6752, 13.6828 \\
\ion{Fe}{xix} &  13.7950 & 6.9 & 7.46e-14 &   7.7 & -0.06 & \ion{Fe}{xix} 13.7315, 13.7458, \ion{Fe}{xx} 13.7670, \ion{Ni}{xix} 13.7790 \\
\ion{Fe}{xvii} &  13.8250 & 6.8 & 3.24e-14 &   5.1 &  0.05 & \ion{Fe}{xix} 13.8390, \ion{Fe}{xx} 13.8430, \ion{Fe}{xvii} 13.8920 \\
\ion{Fe}{xxi} &  14.0080 & 7.0 & 7.22e-14 &   7.6 & -0.17 & \ion{Ni}{xix} 14.0430, 14.0770 \\
\ion{Fe}{xviii} &  14.2080 & 6.9 & 1.06e-13 &   9.3 & -0.13 & \ion{Fe}{xviii} 14.2560, \ion{Fe}{xx} 14.2670 \\
\ion{Fe}{xviii} &  14.3730 & 6.9 & 8.38e-14 &   8.4 &  0.06 & \ion{Fe}{xx} 14.3318, 14.4207, 14.4600 \ion{Fe}{xviii} 14.3430, 14.4250, 14.4392 \\
\ion{Fe}{xviii} &  14.5340 & 6.9 & 5.83e-14 &   7.0 &  0.14 & \ion{Fe}{xviii} 14.4856, 14.5056, 14.5710, 14.6011 \\
\ion{Fe}{xix} &  14.6640 & 6.9 & 2.94e-14 &   5.0 &  0.22 & \ion{Fe}{xviii} 14.6160, 14.6884 \\
\ion{Fe}{xix} &  14.7250 & 6.9 & 1.34e-14 &   3.4 & -0.24 & \ion{Fe}{xviii} 14.7260, 14.7820, \ion{Fe}{xx} 14.7540 \\
\ion{O}{viii} &  14.8205 & 6.5 & 6.04e-14 &   7.3 &  0.34 & \ion{O}{viii} 14.8207, \ion{Fe}{xx} 14.8276, 14.8526, 14.8651, 14.8785, 14.9196, \ion{Fe}{xix} 14.9170, \ion{Fe}{xviii} 14.9241 \\
\ion{Fe}{xvii} &  15.0140 & 6.7 & 2.04e-13 &  13.5 & -0.01 & \ion{Fe}{xix} 15.0790 \\
\ion{Fe}{xvii} &  15.2610 & 6.7 & 5.70e-14 &   7.2 & -0.16 & \ion{O}{viii} 15.1760, 15.1765, \ion{Fe}{xix} 15.1980 \\
\ion{Fe}{xvii} &  15.4530 & 6.7 & 1.93e-14 &   4.2 &  0.19 & \ion{Fe}{xix} 15.4136, \ion{Fe}{xviii} 15.4940, 15.5199, \ion{Fe}{xx} 15.5170 \\
\ion{Fe}{xviii} &  15.6250 & 6.8 & 1.67e-14 &   3.9 & -0.15 &  \\
\ion{Fe}{xviii} &  15.8240 & 6.8 & 3.78e-14 &   6.0 &  0.24 & \ion{Fe}{xviii} 15.8700 \\
\ion{Fe}{xviii} &  16.0710 & 6.8 & 1.50e-13 &  12.0 &  0.06 & \ion{Fe}{xviii} 16.0040, \ion{O}{viii} 16.0055, 16.0067, \ion{Fe}{xix} 16.1100 \\
\ion{Fe}{xviii} &  16.1590 & 6.8 & 1.66e-14 &   4.1 & -0.08 & \ion{Fe}{xvii} 16.2285, \ion{Fe}{xix} 16.2830 \\
\ion{Fe}{xvii} &  16.7800 & 6.7 & 6.92e-14 &   8.3 & -0.03 &  \\
\ion{Fe}{xvii} &  17.0510 & 6.7 & 1.82e-13 &  12.5 &  0.11 & \ion{Fe}{xvii} 17.0960 \\
\ion{O}{viii} &  18.9671 & 6.5 & 2.13e-13 &  14.6 & -0.09 & \ion{O}{viii} 18.9725 \\
\ion{S}{xiv} &  24.2850 & 6.5 & 1.68e-14 &   3.7 &  0.40 & \ion{S}{xiv} 24.2000, 24.2890 \\
\ion{N}{vii} &  24.7792 & 6.3 & 5.34e-14 &   6.6 &  0.00 & \ion{N}{vii} 24.7846 \\
\ion{C}{vi} &  28.4652 & 6.2 & 2.11e-14 &   4.3 &  0.00 & \ion{C}{vi} 28.4663 \\
\ion{S}{xiv} &  30.4270 & 6.5 & 1.66e-14 &   3.5 & -0.32 & \ion{S}{xiv} 30.4690 \\
No id. &  36.3980 &  & 4.30e-14 &   6.0 & \ldots & (\ion{Ne}{x} 12.132, 3$^{\rm rd}$ order) \\
\ion{Si}{xii} &  44.1650 & 6.3 & 9.40e-15 &   4.5 & -0.12 & \ion{Si}{xii} 44.0190, 44.1780 \\
\ion{Fe}{xvii} &  50.6861 & 6.8 & 3.99e-15 &   3.7 &  0.32 & \ion{Fe}{xvi} 50.5550, \ion{Fe}{xvii} 50.8544 \\
No id. &  56.9000 &  & 1.07e-14 &   4.1 & \ldots & (\ion{O}{viii} 18.97, 3$^{\rm rd}$ order) \\
\ion{Fe}{xviii} &  93.9230 & 6.8 & 2.87e-14 &   8.6 & -0.29 & \ion{Fe}{xx} 93.7800 \\
\ion{Fe}{xix} & 101.5500 & 6.9 & 1.30e-14 &   3.7 & -0.21 &  \\
\ion{Fe}{xxi} & 102.2200 & 7.0 & 1.01e-14 &   3.3 & -0.43 &  \\
\ion{Fe}{xix} & 108.3700 & 6.9 & 3.99e-14 &   6.4 & -0.15 &  \\
\ion{Fe}{xxii} & 117.1700 & 7.1 & 1.92e-14 &   4.4 & -0.54 &  \\
\ion{Fe}{xxi} & 128.7300 & 7.0 & 4.25e-14 &   4.9 & -0.31 &  \\
\ion{Fe}{xx} & 132.8500 & 7.0 & 7.64e-14 &   6.6 & -0.38 & \ion{Fe}{xxiii} 132.8500 \\
\ion{Fe}{xxii} & 135.7800 & 7.1 & 5.89e-14 &   5.8 &  0.14 &  \\
\hline
\end{tabular}

{$^a$ Line fluxes (in erg cm$^{-2}$ s$^{-1}$) 
  measured in Chandra/LETG AY Cet summed spectra, and corrected by the
  ISM absorption. 
  log $T_{\mathrm {max}}$ indicates the maximum
  temperature (K) of formation of the line (unweighted by the
  EMD). ``Ratio'' is the log~($F_{\mathrm {obs}}$/$F_{\mathrm {pred}}$) 
  of the line. 
  Blends amounting to more than 5\% of the total flux for each line are
  indicated. For some lines not identified in APED, a tentative identification as 3$^{\rm rd}$ order emission of intense lines is suggested in the ``Blends'' column.}
\end{scriptsize}
\end{table*}
}

%% file: abund2paper.bbl
\begin{thebibliography}{47}
\expandafter\ifx\csname natexlab\endcsname\relax\def\natexlab#1{#1}\fi

\bibitem[{{Affer} {et~al.}(2005){Affer}, {Micela}, {Morel}, {Sanz-Forcada}, \&
  {Favata}}]{aff05}
{Affer}, L., {Micela}, G., {Morel}, T., {Sanz-Forcada}, J., \& {Favata}, F.
  2005, \aap, 433, 647

\bibitem[{{Anders} \& {Grevesse}(1989)}]{anders}
{Anders}, E. \& {Grevesse}, N. 1989, \gca, 53, 197

\bibitem[{{Argiroffi} {et~al.}(2003){Argiroffi}, {Maggio}, \& {Peres}}]{arg03}
{Argiroffi}, C., {Maggio}, A., \& {Peres}, G. 2003, \aap, 404, 1033

\bibitem[{{Asplund} {et~al.}(2005){Asplund}, {Grevesse}, \& {Sauval}}]{asp05}
{Asplund}, M., {Grevesse}, N., \& {Sauval}, A.~J. 2005, in ASP Conf. Series,
  Vol. 336, Cosmic Abundances as Records of Stellar Evolution and
  Nucleosynthesis, ed. T.~G. {Barnes}, III \& F.~N. {Bash}, 25

\bibitem[{{Audard} {et~al.}(2003){Audard}, {G{\"u}del}, {Sres}, {Raassen}, \&
  {Mewe}}]{aud03}
{Audard}, M., {G{\"u}del}, M., {Sres}, A., {Raassen}, A.~J.~J., \& {Mewe}, R.
  2003, \aap, 398, 1137

\bibitem[{{Audard} {et~al.}(2004){Audard}, {Telleschi}, {G{\"u}del}, {Skinner},
  {Pallavicini}, \& {Mitra-Kraev}}]{aud04}
{Audard}, M., {Telleschi}, A., {G{\"u}del}, M., {et~al.} 2004, \apj, 617, 531

\bibitem[{{Ball} {et~al.}(2005){Ball}, {Drake}, {Lin}, {Kashyap}, {Laming}, \&
  {Garc{\'{\i}}a-Alvarez}}]{bal05}
{Ball}, B., {Drake}, J.~J., {Lin}, L., {et~al.} 2005, \apj, 634, 1336

\bibitem[{{Brickhouse} {et~al.}(2000){Brickhouse}, {Dupree}, {Edgar},
  {Liedahl}, {Drake}, {White}, \& {Singh}}]{bri00}
{Brickhouse}, N.~S., {Dupree}, A.~K., {Edgar}, R.~J., {et~al.} 2000, \apj, 530,
  387

\bibitem[{{Brinkman} {et~al.}(2001){Brinkman}, {Behar}, {G{\"u}del}, {Audard},
  {den Boggende}, {Branduardi-Raymont}, {Cottam}, {Erd}, {den Herder},
  {Jansen}, {Kaastra}, {Kahn}, {Mewe}, {Paerels}, {Peterson}, {Rasmussen},
  {Sakelliou}, \& {de Vries}}]{bri01}
{Brinkman}, A.~C., {Behar}, E., {G{\"u}del}, M., {et~al.} 2001, \aap, 365, L324

\bibitem[{{De Pontieu} {et~al.}(2007){De Pontieu}, {McIntosh}, {Carlsson},
  {Hansteen}, {Tarbell}, {Schrijver}, {Title}, {Shine}, {Tsuneta}, {Katsukawa},
  {Ichimoto}, {Suematsu}, {Shimizu}, \& {Nagata}}]{dep07}
{De Pontieu}, B., {McIntosh}, S.~W., {Carlsson}, M., {et~al.} 2007, Science,
  318, 1574

\bibitem[{{den Herder} {et~al.}(2001){den Herder}, {Brinkman}, {Kahn},
  {Branduardi-Raymont}, {Thomsen}, {Aarts}, {Audard}, {Bixler}, {den Boggende},
  {Cottam}, {Decker}, {Dubbeldam}, {Erd}, {Goulooze}, {G{\" u}del},
  {Guttridge}, {Hailey}, {Janabi}, {Kaastra}, {de Korte}, {van Leeuwen},
  {Mauche}, {McCalden}, {Mewe}, {Naber}, {Paerels}, {Peterson}, {Rasmussen},
  {Rees}, {Sakelliou}, {Sako}, {Spodek}, {Stern}, {Tamura}, {Tandy}, {de
  Vries}, {Welch}, \& {Zehnder}}]{denher01}
{den Herder}, J.~W., {Brinkman}, A.~C., {Kahn}, S.~M., {et~al.} 2001, \aap,
  365, L7

\bibitem[{{Drake} {et~al.}(1997){Drake}, {Laming}, \& {Widing}}]{dra97}
{Drake}, J.~J., {Laming}, J.~M., \& {Widing}, K.~G. 1997, \apj, 478, 403

\bibitem[{{Drake} \& {Testa}(2005)}]{dra05}
{Drake}, J.~J. \& {Testa}, P. 2005, \nat, 436, 525

\bibitem[{{Erd{\'e}lyi} \& {Fedun}(2007)}]{erd07}
{Erd{\'e}lyi}, R. \& {Fedun}, V. 2007, Science, 318, 1572

\bibitem[{{Favata} \& {Micela}(2003)}]{fav03}
{Favata}, F. \& {Micela}, G. 2003, Space Science Reviews, 108, 577

\bibitem[{{Feldman} \& {Laming}(2000)}]{feld00}
{Feldman}, U. \& {Laming}, J.~M. 2000, \physscr, 61, 222

\bibitem[{{Garc{\'{\i}}a-Alvarez} {et~al.}(2008){Garc{\'{\i}}a-Alvarez},
  {Drake}, {Kashyap}, {Lin}, \& {Ball}}]{gar08}
{Garc{\'{\i}}a-Alvarez}, D., {Drake}, J.~J., {Kashyap}, V.~L., {Lin}, L., \&
  {Ball}, B. 2008, \apj, 679, 1509

\bibitem[{{Garc{\'{\i}}a-Alvarez} {et~al.}(2005){Garc{\'{\i}}a-Alvarez},
  {Drake}, {Lin}, {Kashyap}, \& {Ball}}]{gar05}
{Garc{\'{\i}}a-Alvarez}, D., {Drake}, J.~J., {Lin}, L., {Kashyap}, V.~L., \&
  {Ball}, B. 2005, \apj, 621, 1009

\bibitem[{{G{\"u}del}(2004)}]{gud04}
{G{\"u}del}, M. 2004, \aapr, 12, 71

\bibitem[{{Huenemoerder} {et~al.}(2003){Huenemoerder}, {Canizares}, {Drake}, \&
  {Sanz-Forcada}}]{hue03}
{Huenemoerder}, D.~P., {Canizares}, C.~R., {Drake}, J.~J., \& {Sanz-Forcada},
  J. 2003, \apj, 595, 1131

\bibitem[{{Huenemoerder} {et~al.}(2001){Huenemoerder}, {Canizares}, \&
  {Schulz}}]{hue01}
{Huenemoerder}, D.~P., {Canizares}, C.~R., \& {Schulz}, N.~S. 2001, \apj, 559,
  1135

\bibitem[{{Kurucz}(1993)}]{kur93}
{Kurucz}, R.~L. 1993, {SYNTHE spectrum synthesis programs and line data}, ed.
  R.~L. {Kurucz}

\bibitem[{{Laming}(2004)}]{lam04}
{Laming}, J.~M. 2004, \apj, 614, 1063

\bibitem[{{Laming} {et~al.}(1995){Laming}, {Drake}, \& {Widing}}]{lam95}
{Laming}, J.~M., {Drake}, J.~J., \& {Widing}, K.~G. 1995, \apj, 443, 416

\bibitem[{{Laming} {et~al.}(1996){Laming}, {Drake}, \& {Widing}}]{lam96}
{Laming}, J.~M., {Drake}, J.~J., \& {Widing}, K.~G. 1996, \apj, 462, 948

\bibitem[{{Massarotti} {et~al.}(2008){Massarotti}, {Latham}, {Stefanik}, \&
  {Fogel}}]{mas08}
{Massarotti}, A., {Latham}, D.~W., {Stefanik}, R.~P., \& {Fogel}, J. 2008, \aj,
  135, 209

\bibitem[{{McWilliam}(1990)}]{mcw90}
{McWilliam}, A. 1990, \apjs, 74, 1075

\bibitem[{{Montes} {et~al.}(1997){Montes}, {Fernandez-Figueroa}, {de Castro},
  \& {Sanz-Forcada}}]{mon97}
{Montes}, D., {Fernandez-Figueroa}, M.~J., {de Castro}, E., \& {Sanz-Forcada},
  J. 1997, \aaps, 125, 263

\bibitem[{{Morel} {et~al.}(2003){Morel}, {Micela}, {Favata}, {Katz}, \&
  {Pillitteri}}]{mor03}
{Morel}, T., {Micela}, G., {Favata}, F., {Katz}, D., \& {Pillitteri}, I. 2003,
  \aap, 412, 495

\bibitem[{{Ness} \& {Jordan}(2008)}]{ness08}
{Ness}, J.-U. \& {Jordan}, C. 2008, \mnras, 385, 1691

\bibitem[{{Nordstr{\"o}m} {et~al.}(2004){Nordstr{\"o}m}, {Mayor}, {Andersen},
  {Holmberg}, {Pont}, {J{\o}rgensen}, {Olsen}, {Udry}, \& {Mowlavi}}]{nor04}
{Nordstr{\"o}m}, B., {Mayor}, M., {Andersen}, J., {et~al.} 2004, \aap, 418, 989

\bibitem[{{Ottmann} {et~al.}(1998){Ottmann}, {Pfeiffer}, \& {Gehren}}]{ott98}
{Ottmann}, R., {Pfeiffer}, M.~J., \& {Gehren}, T. 1998, \aap, 338, 661

\bibitem[{{Raassen} {et~al.}(2002){Raassen}, {Mewe}, {Audard}, {G{\" u}del},
  {Behar}, {Kaastra}, {van der Meer}, {Foley}, \& {Ness}}]{raa02}
{Raassen}, A.~J.~J., {Mewe}, R., {Audard}, M., {et~al.} 2002, \aap, 389, 228

\bibitem[{{Raassen} {et~al.}(2003){Raassen}, {Ness}, {Mewe}, {van der Meer},
  {Burwitz}, \& {Kaastra}}]{raa03}
{Raassen}, A.~J.~J., {Ness}, J.-U., {Mewe}, R., {et~al.} 2003, \aap, 400, 671

\bibitem[{{Randich} {et~al.}(1993){Randich}, {Gratton}, \&
  {Pallavicini}}]{ran93}
{Randich}, S., {Gratton}, R., \& {Pallavicini}, R. 1993, \aap, 273, 194

\bibitem[{{Sanz-Forcada} {et~al.}(2003{\natexlab{a}}){Sanz-Forcada},
  {Brickhouse}, \& {Dupree}}]{sanz03}
{Sanz-Forcada}, J., {Brickhouse}, N.~S., \& {Dupree}, A.~K. 2003{\natexlab{a}},
  \apjs, 145, 147

\bibitem[{{Sanz-Forcada} {et~al.}(2004){Sanz-Forcada}, {Favata}, \&
  {Micela}}]{sanz04}
{Sanz-Forcada}, J., {Favata}, F., \& {Micela}, G. 2004, \aap, 416, 281

\bibitem[{{Sanz-Forcada} {et~al.}(2003{\natexlab{b}}){Sanz-Forcada}, {Maggio},
  \& {Micela}}]{sanz03b}
{Sanz-Forcada}, J., {Maggio}, A., \& {Micela}, G. 2003{\natexlab{b}}, \aap,
  408, 1087

\bibitem[{{Schmelz} {et~al.}(2005){Schmelz}, {Nasraoui}, {Roames}, {Lippner},
  \& {Garst}}]{sch05}
{Schmelz}, J.~T., {Nasraoui}, K., {Roames}, J.~K., {Lippner}, L.~A., \&
  {Garst}, J.~W. 2005, \apjl, 634, L197

\bibitem[{{Schmitt} {et~al.}(1996){Schmitt}, {Stern}, {Drake}, \&
  {Kuerster}}]{sch96}
{Schmitt}, J.~H.~M.~M., {Stern}, R.~A., {Drake}, J.~J., \& {Kuerster}, M. 1996,
  \apj, 464, 898

\bibitem[{{Shan} {et~al.}(2006){Shan}, {Liu}, \& {Gu}}]{sha06}
{Shan}, H., {Liu}, X., \& {Gu}, S. 2006, New Astronomy, 11, 287

\bibitem[{{Smith} {et~al.}(2001){Smith}, {Brickhouse}, {Liedahl}, \&
  {Raymond}}]{aped}
{Smith}, R.~K., {Brickhouse}, N.~S., {Liedahl}, D.~A., \& {Raymond}, J.~C.
  2001, \apjl, 556, L91

\bibitem[{{Sneden}(1973)}]{sne73}
{Sneden}, C.~A. 1973, PhD thesis, The University of Texas at Austin)

\bibitem[{{Suh} {et~al.}(2005){Suh}, {Audard}, {G{\"u}del}, \&
  {Paerels}}]{suh05}
{Suh}, J.~A., {Audard}, M., {G{\"u}del}, M., \& {Paerels}, F.~B.~S. 2005, \apj,
  630, 1074

\bibitem[{{Telleschi} {et~al.}(2005){Telleschi}, {G{\"u}del}, {Briggs},
  {Audard}, {Ness}, \& {Skinner}}]{tel05}
{Telleschi}, A., {G{\"u}del}, M., {Briggs}, K., {et~al.} 2005, \apj, 622, 653

\bibitem[{{Weisskopf} {et~al.}(2002){Weisskopf}, {Brinkman}, {Canizares},
  {Garmire}, {Murray}, \& {Van Speybroeck}}]{wei02}
{Weisskopf}, M.~C., {Brinkman}, B., {Canizares}, C., {et~al.} 2002, \pasp, 114,
  1

\bibitem[{{Wood} \& {Linsky}(2006)}]{woo06}
{Wood}, B.~E. \& {Linsky}, J.~L. 2006, \apj, 643, 444

\end{thebibliography}
